\documentclass[aps,prd,reprint,amsmath,floatfix,showpacs]{revtex4-1}

\usepackage{graphicx} 

\begin{document}

\preprint{
DESY-13-153  
}

\title{
{\hfill DESY-13-153} \\[11mm] 
 Precise Calculation of the Dilepton 
 Invariant-Mass Spectrum \\ and the Decay Rate  
 in $B^\pm \to \pi^\pm \mu^+ \mu^-$ in the SM
}



\author{Ahmed Ali}
\email[]{ahmed.ali@desy.de}
\affiliation{
 Theory Group,
 Deutsches Elektronen-Synchrotron DESY,
 D-22603 Hamburg, FRG 
}

\author{Alexander Ya.~Parkhomenko}
\email[]{parkh@uniyar.ac.ru}
\affiliation{
 Department of Theoretical Physics,
 P.\,G.~Demidov Yaroslavl State University, 
 Sovietskaya 14, 150000 Yaroslavl, Russia 
}

\author{Aleksey V.~Rusov}
\email[]{rusov@uniyar.ac.ru}
\affiliation{
 Department of Theoretical Physics,
 P.\,G.~Demidov Yaroslavl State University, 
 Sovietskaya 14, 150000 Yaroslavl, Russia   
}  
\affiliation{
 Department of Physics,
 A.\,F.~Mozhaisky Military Space Academy (Yaroslavl Branch), 
 Moskovsky Prospect 28, 150001 Yaroslavl, Russia 
}


\date{\today}

\begin{abstract} 
We present a precise calculation of the dilepton invariant-mass
spectrum and the decay rate for $B^\pm \to \pi^\pm \ell^+ \ell^-$ 
($\ell^\pm = e^\pm, \mu^\pm $) in the Standard Model (SM) based 
on the effective Hamiltonian approach for the $b \to d \ell^+ \ell^-$ 
transitions. With the Wilson coefficients already known in the 
next-to-next-to-leading logarithmic (NNLL) accuracy, the remaining 
theoretical uncertainty in the short-distance contribution resides 
in the form factors~$f_+ (q^2)$, $f_0 (q^2)$ and~$f_T (q^2)$. 
Of these, $f_+ (q^2)$ is well measured in the charged-current
semileptonic decays $B \to \pi \ell \nu_\ell$ and we use 
the $B$-factory data to parametrize it. The corresponding 
form factors for the $B \to K$ transitions have been calculated 
in the Lattice-QCD approach for large-$q^2$ and extrapolated 
to the entire $q^2$-region using the so-called $z$-expansion. 
Using an $SU (3)_F$-breaking Ansatz, we calculate the $B \to \pi$ 
tensor form factor, which is consistent with the recently 
reported lattice $B \to \pi$ analysis obtained at large~$q^2$. 
The prediction for the total branching fraction 
${\cal B} (B^\pm \to \pi^\pm \mu^+ \mu^-) = 
 (1.88 ^{+0.32}_{-0.21}) \times 10^{-8}$ is in good agreement 
with the experimental value obtained by the LHCb Collaboration. 
In the low $q^2$-region, heavy-quark symmetry (HQS) relates 
the three form factors with each other. Accounting for the 
leading-order symmetry-breaking effects, and using data from 
the charged-current process $B \to \pi \ell \nu_\ell$ 
to determine $f_+ (q^2)$, we calculate the dilepton 
invariant-mass distribution in the low $q^2$-region 
in the $B^\pm \to \pi^\pm \ell^+ \ell^-$ decay. 
This provides a model-independent and precise calculation 
of the partial branching ratio for this decay.
\end{abstract}


\maketitle


\section{\label{sec:introduction} 
Introduction 
}

Recently, the LHCb Collaboration has reported the first 
observation of the $B^\pm \to \pi^\pm \mu^+ \mu^-$ decay, 
using 1.0~fb$^{-1}$ 
integrated luminosity in proton-proton collisions 
at the Large Hadron Collider (LHC)
at $\sqrt s = 7$~TeV~\cite{LHCb:2012de}.
Unlike the $b \to s \, \ell^+ \ell^-$ transitions, which 
have been studied at the $B$-factories and hadron colliders 
in a number of decays, such as $B \to (K,\, K^*) \, \ell^+ \ell^-$ 
and $B_s \to \phi \, \ell^+ \ell^-$~\cite{Amhis:2012bh}, 
the $B^\pm \to \pi^\pm \, \mu^+ \mu^-$ decay is the first 
$b \to d \, \ell^+ \ell^-$ transition measured so far. 
Phenomenological analysis of this process, under controlled 
theoretical errors, will provide us independent information 
concerning the $b \to d$ Flavor-Changing-Neutral-Current
(FCNC) transitions in the $B$-meson sector. Hence,  
$B^\pm \to \pi^\pm \mu^+ \mu^-$ decay is potentially an important 
input in the precision tests of the~SM in the flavor sector 
and, by the same token, also in searches for physics beyond it.

The measured branching ratio 
${\cal B} (B^+ \to \pi^+ \mu^+ \mu^-) = 
[ 2.3 \pm 0.6 ({\rm stat}) \pm 0.1 ({\rm syst}) ]  
\times 10^{-8}$~\cite{LHCb:2012de} is in good agreement 
with the SM expected rate~\cite{Wang:2007sp}, which, 
however, like a number of other estimates in the 
literature~\cite{Song:2008zzc,Wang:2012ab}, is based 
on model-dependent input for the $B \to \pi$ form factors. 
The Light-Cone Sum Rules (LCSR) approach (see, for example, 
Refs.~\cite{Ball:2004ye} and~\cite{Duplancic:2008ix}) 
is certainly helpful in the low $q^2$-region and has been used 
in the current phenomenological analysis of the data~\cite{LHCb:2012de}. 
However, theoretical accuracy of the LCSR-based form factors 
is limited due to the dependence on numerous input parameters 
and wave-function models. Hence, it is very desirable to calculate 
the form factors from first principles, such as the Lattice QCD, 
which have their own range of validity restricted by the recoil 
energy (here, the energy~$E_\pi$ of the $\pi$-meson), as the 
discretization errors become large with increasing~$E_\pi$. 
With improved lattice technology, one can use the lattice 
form factors to predict the decay rates in the $B \to \pi$ and 
$B \to K$ transitions (as well as in other heavy-to-light meson 
transitions) in the low-recoil region, where the lattice results 
apply without any extrapolation, in a model-independent manner. 
At present, the dimuon invariant mass distribution 
in the $B^+ \to \pi^+ \, \mu^+ \mu^-$ decay is not at hand 
and only the integrated branching ratio is known. We combine 
the lattice input with other phenomenologically robust
approaches to calculate the dilepton invariant-mass spectrum 
in the entire $q^2$-region to compute the corresponding integrated 
decay rates for comparison with the data~\cite{LHCb:2012de}.
Our framework makes use of the methods based on the heavy-quark 
symmetry (HQS) in the large-recoil region, data from the 
$B$-factory experiments on the charged-current processes 
\footnote{The charge conjugation is implicit in this paper.}  
$B^0 \to \pi^- \ell^+ \nu_\ell$ and $B^+ \to \pi^0 \ell^+ \nu_\ell$
to determine one of the form factors, $f_+ (q^2)$, and 
the available lattice results on the $B \to \pi$ and 
$B \to K$ transition form factors in the low-recoil region. 

We recall that the decay $B^\pm \to \pi^\pm \, \ell^+ \ell^-$ 
involves three form factors, two of which, $f_+ (q^2)$ and 
$f_0 (q^2)$, characterize the hadronic $B \to \pi$ matrix 
element of the vector current 
$J_V^\mu (x) = \bar b (x) \gamma^\mu d (x)$, 
and the third, $f_T (q^2)$, enters 
in the corresponding matrix element of the tensor current 
$J_T^\mu (x) = \bar b (x) \sigma^{\mu\nu} q_\nu d (x)$, 
where $q^\mu = p_B^\mu - p_\pi^\mu$ is the momentum transferred 
to the lepton pair $\ell^+ \ell^-$ (see Eqs.~(\ref{eq:ME-vector}) 
and~(\ref{eq:ME-tensor}) below). Using the isospin symmetry, 
the first two form factors are the same as the ones encountered 
in the charged-current processes $B^+ \to \pi^0 \ell^+ \nu_\ell$ 
and $B^0 \to \pi^- \ell^+ \nu_\ell$. Of these, the contribution to 
the decay rate proportional to $f_0 (q^2)$ is strongly suppressed 
by the mass ratio $m_\ell^2/m_B^2$ (for $\ell = e, \, \mu$). 
The form factor $f_+ (q^2)$ has been well measured 
(modulo~$|V_{ub}|$) in the entire $q^2$-range by the 
BaBar~\cite{delAmoSanchez:2010af,Lees:2012mq} and 
Belle~\cite{Ha:2010rf,Sibidanov:2013rkk} collaborations. 
We have undertaken a chi-squared fit of these data, using 
four popular form-factor parametrizations of $f_+ (q^2)$: 
(i) the Becirevic-Kaidalov (BK) parametrization~\cite{Becirevic:1999kt},
(ii) the Ball-Zwicky (BZ) parametrization~\cite{Ball:2004ye}, 
(iii) the Boyd-Grinstein-Lebed (BGL) parametrization~\cite{Boyd:1994tt}, 
and (iv) the Bourrely-Caprini-Lellouch (BCL) 
parametrization~\cite{Bourrely:2008za}. 
All these parametrizations yield good fits measured 
in terms of $\chi^2_{\rm min}/{\rm ndf}$, 
where ndf is the number of degrees of freedom 
(see Table~\ref{tab:chisqvalues}). 
However, factoring in theoretical arguments based on the 
Soft-Collinear Effective Theory (SCET)~\cite{Becher:2005bg}, 
and preference of the Lattice-QCD-based analysis of the form 
factors $f_+ (q^2)$, $f_0 (q^2)$, and $f_T (q^2)$ in terms 
of the so-called $z$-expansion, and a variation thereof 
(see Ref.~\cite{Zhou:2013uu} for a recent summary of the lattice 
heavy-to-light form factors), we use the BGL-parametrization 
as our preferred choice for the extraction of $f_+ (q^2)$ 
from the $B \to \pi \ell \nu_\ell$ data.
It should be noted that our analysis for the extraction 
of $f_+ (q^2)$ is model independent as it is based on the 
complete set of experimental data. Meanwhile, there are 
also several theoretical non-perturbative methods which 
allows one to determine this form factor but usually 
in a limited $q^2$-range, for example the
 LCSRs~\cite{Duplancic:2008ix,Khodjamirian:2011ub,Bharucha:2012wy} 
and the $k_T$-factorization approach~\cite{Li:2012nk}, which are
often invoked in estimating 
the vector $B \to \pi$ transition form factor.

In order to determine the other two form factors, $f_0 (q^2)$ 
and $f_T (q^2)$, in the entire $q^2$-domain, we proceed 
as follows: Lattice QCD provides them in the high-$q^2$ region. 
A number of dedicated lattice-based studies of the heavy-to-light 
form factors are available in the literature. 
In particular, calculations of the form factors 
in the $B \to (K, \, K^*) \, \ell^+ \ell^-$ decays, 
based on the $(2 + 1)$-flavor gauge configurations generated 
by the MILC Collaboration~\cite{Bazavov:2009bb}, have been 
undertaken by the FNAL/MILC~\cite{Zhou:2011be,Zhou:2012dm}, 
HPQCD~\cite{Bouchard:2013mia,Bouchard:2013eph} and the 
Cambridge/Edinburgh~\cite{Liu:2009dj,Liu:2011raa} Lattice groups. 
We make use of the $B \to K$ lattice results, combining them 
with an Ansatz on the $SU (3)_F$-symmetry breaking to determine 
the $f_T (q^2)$ form factor for the $B \to \pi$ transition. 
Very recently, new results on the $B \to \pi$ form factors, 
in particular the first preliminary results on the tensor 
form factor $f_T^{B\pi} (q^2)$, from the lattice simulations 
have also become available~\cite{Bouchard:2013zda,Du:2013kea}. 
While the analysis presented in Ref.~\cite{Du:2013kea} 
by the FermiLab Lattice and MILC Collaborations is still blinded 
with an unknown off-set factor, promised to be disclosed 
when the final results are presented, we use the available 
results on the $f_T^{B K} (q^2)$ form factor by the HPQCD 
Collaboration~\cite{Bouchard:2013mia,Bouchard:2013eph} 
as input in the high-$q^2$ region to constrain our Ansatz 
on the $SU(3)_F$-symmetry breaking. Thus, combining the 
extraction of $f_+ (q^2)$ from the $B \to \pi \ell \nu_\ell$ 
data, the Lattice-QCD data on $f_T (q^2)$ for the large-$q^2$ 
domain, and the BGL-like parametrization~\cite{Boyd:1994tt} 
in the form of $z$-expansion to extrapolate this form factor to 
the lower $q^2$-range, we obtain the following branching ratio:
\begin{equation} 
{\cal B} (B^+ \to \pi^+ \mu^+ \mu^-) = 
(1.88^{+0.32}_{-0.21} ) \times 10^{-8}~,  
\label{eq:Ourbr}
\end{equation}
which has a combined accuracy of about $\pm 15$\%, 
taking into account also the uncertainties 
in the CKM matrix elements, for which we have 
used the values obtained from the fits of 
the CKM unitarity triangle~\cite{Beringer:1900zz}.  
This result is in agreement (within large experimental 
errors) with the experimental value reported recently 
by the LHCb Collaboration~\cite{LHCb:2012de}: 
\begin{equation} 
{\cal B} (B^+ \to \pi^+ \mu^+ \mu^-) = \left ( 
2.3 \pm 0.6 ({\rm stat.}) \pm 0.1({\rm syst.}) 
\right ) \times 10^{-8}.  
\label{eq:LHCb2012}
\end{equation}
As the lattice calculations of the $B \to \pi$ form factors 
become robust and the dilepton invariant-mass spectrum 
in $B^+ \to \pi^+ \mu^+\mu^-$ is measured, one can undertake 
a completely quantitative fit of the data in the~SM taking 
into account correlations in the lattice calculations and data.

In the SM, the $b \to d \, \ell^+ \ell^-$ transition 
is suppressed essentially by the factor $|V_{td}/V_{ts}|$ 
relative to the $b \to s \, \ell^+ \ell^-$ transition. 
In terms of exclusive decays, first measurement of the ratio
${\cal B} (B^+ \to \pi^+ \ell^+ \ell^-)/
 {\cal B} (B^+ \to K^+ \ell^+ \ell^-)$ 
has been reported by the LHCb Collaboration~\cite{LHCb:2012de}:   
\begin{equation}
\frac{{\cal B} (B^+ \to \pi^+ \mu^+ \mu^-)}
     {{\cal B} (B^+ \to K^+ \mu^+ \mu^-)} 
= 0.053 \pm 0.014 ({\rm stat.}) \pm 0.001 ({\rm syst.})~.
\label{eq:brpik}
\end{equation}
In the SM, this ratio can be expressed as follows:
\begin{equation} 
\frac{{\cal B} (B^+ \to \pi^+ \mu^+ \mu^-)}
     {{\cal B} (B^+ \to K^+ \mu^+ \mu^-)} = 
\left | \frac{V_{td}}{V_{ts}} \right |^2 \, 
F^{\pi/K}_{\rm tot}~, 
\label{eq:pi/K-ratio} 
\end{equation}
where $F^{\pi/K}_{\rm tot}$ is the ratio resulting 
from the convolution of the form factors and 
the $q^2$-dependent effective Wilson coefficients. 
Using $F^{\pi/K}_{\rm tot} = 0.87$, and neglecting 
the errors on this quantity, LHCb has determined 
the ratio of the CKM matrix elements, yielding
$|V_{td}/V_{ts}| = 0.266 \pm 0.035 ({\rm stat.}) 
       \pm 0.003 ({\rm syst.})$~\cite{LHCb:2012de}.
At present this method is not competitive with other 
determinations of~$|V_{td}/V_{ts}|$, such as from 
the $B_{(s)} - \bar B_{(s)}$ mixings~\cite{Amhis:2012bh}, 
but with greatly improved statistical error, anticipated 
at the LHC and Super-$B$ factory experiments, this
would become a valuable and independent constraint 
on the CKM matrix. A reliable estimate of the quantity
$F^{\pi/K}_{\rm tot}$ is also required. In particular, 
we expect that the error on the corresponding quantity, 
$F^{\pi/K}_{\rm HQS} (q^2 \leq q_0^2)$, denoting the 
ratio of the partial branching ratios restricted to the 
low-$q^2$ domain, can be largely reduced with the help 
of the heavy-quark symmetry. We hope to return to 
improved theoretical estimates of $F^{\pi/K}_{\rm tot}$ and 
$F^{\pi/K}_{\rm HQS} (q^2 \leq q_0^2)$ in a future publication.

In the large-recoil limit, the form factors in the 
$B \to (\pi, \, \rho, \, \omega)$ and $B \to (K, \, K^*)$ 
transitions obey the heavy-quark symmetry, reducing 
the number of independent form factors~\cite{Charles:1998dr}.
In particular, the $B \to \pi$ form factors $f_0 (q^2)$ 
and $f_T (q^2)$ are related to $f_+ (q^2)$ in the HQS limit 
(see Eqs.~(\ref{eq:fp-f0-relation}) and~(\ref{eq:fp-fT-relation}) 
below). Taking into account the leading-order symmetry-breaking 
corrections, these relations get modified~\cite{Beneke:2001at}, 
bringing in their wake a dependence on the QCD coupling 
constant $\alpha_s (\mu_h)$ and $\alpha_s (\mu_{hc})$, 
where the hard scale $\mu_h \simeq m_b$ and the intermediate 
(or hard-collinear) scale $\mu_{hc}= \sqrt{m_b \Lambda}$,
with $\Lambda \simeq 0.5$~GeV, reflect the multi-scale 
nature of this problem. In addition, a non-perturbative 
quantity $\Delta F_\pi$, which involves the leptonic decay 
constants~$f_B$ and~$f_\pi$ and the first inverse moments 
of the leading-twist light-cone distribution amplitudes 
(LCDAs) of the $B$- and $\pi$-meson also enters 
(see Eqs.~(\ref{eq:f0-fp-rel}) and~(\ref{eq:fT-fp-rel}) below). 
We have used the HQS-based approach to determine the $f_T (q^2)$ 
form factor in terms of the measured $f_+ (q^2)$ form factor from 
the semileptonic $B \to \pi \ell \nu_\ell$ data, discussed above. 
This provides a model-independent determination of the 
dilepton invariant-mass distribution in the low-$q^2$ region.

Leaving uncertainties from the form factors aside, 
the other main problem from the theoretical point of view 
in the $b \to d \, \ell^+ \ell^-$ transitions is the 
so-called long-distance contributions, which are dominated 
by the~$\bar c c$ and~$\bar u u$ resonant states which 
show up as charmonia ($J/\psi$, $\psi (2S)$,~$\ldots$)  
and light vector ($\rho$ and $\omega$) mesons, respectively.
Only model-dependent descriptions (in a Breit-Wigner form) 
of such long-distance effects are known at present, which 
compromise the precision in the theoretical predictions 
of the total branching fractions. Excluding the resonance-dominated 
regions from the dilepton invariant-mass distributions 
is therefore the preferred way to compare data and theory.
With this in mind, we calculate the following partially 
integrated branching ratio
\begin{eqnarray}
{\cal B} (B^+ \to \pi^+ \mu^+ \mu^-; \, 
         1 \,{\rm GeV}^2 \leq q^2 \leq 8 \,{\rm GeV}^2) 
\label{eq:Br-1-8-inrod} \\ 
= \left ( 0.57^{+0.07}_{-0.05} \right ) \times 10^{-8}~,
\nonumber 
\end{eqnarray}
where the lower and upper $q^2$-value boundaries 
are chosen to remove the light-vector ($\rho$- 
and $\omega$-mesons) and charmonium-resonant regions. 
However, with the product branching ratios~\cite{Beringer:1900zz}: 
${\cal B} (B^+ \to \rho^0 \pi^+) \times 
 {\cal B} (\rho^0 \to \mu^+ \mu^-) = 
 \left ( 3.78 \pm 0.59 \right ) \times 10^{-10}$ 
and 
${\cal B} (B^+ \to \omega \pi^+) \times 
 {\cal B} (\omega \to \mu^+ \mu^-) = 
 \left ( 6.2 \pm 2.2 \right ) \times 10^{-10}$,
the long-distance effects in the low-$q^2$ region 
are numerically not important. 

Due to the small branching ratio, it will be a while 
before the entire dimuon invariant mass is completely 
measured in the $B^+ \to \pi^+ \, \mu^+ \mu^-$ decay. 
Anticipating this, and following similar procedures 
adopted in the analysis of the data in the 
$B \to (K, \, K^*) \, \ell^+ \ell^-$ 
decays~\cite{Bobeth:2012vn,Hambrock:2012dg}
we present here results for the partial branching ratios
$d{\cal B} (B^+ \to \pi^+ \mu^+ \mu^-)/dq^2$, binned over 
specified ranges $[q^2_{\rm min}, q^2_{\rm max}]$ in eight 
$q^2$-intervals. They would allow the experiments to check 
the short-distance (renormalization-improved perturbative)
part of the SM contribution in the $b \to d \, \ell^+ \ell^-$ 
transitions precisely.


This paper is organized as follows: 
In Section~\ref{sec:B-pi-mu-mu-basis}, 
we present the dilepton invariant-mass spectrum 
$d{\cal B} (B^+ \to \pi^+ \, \mu^+ \mu^-)/dq^2$ 
in the effective weak Hamiltonian approach based on the~SM 
and the numerical values of the effective Wilson coefficients. 
Section~\ref{sec:FF-parametrizations} 
is devoted to the four popular parameterizations 
of the vector, scalar and tensor form factors.  
Section~\ref{sec:FF-fits} describes the fits of the 
semileptonic data on the $B \to \pi \ell \nu_\ell$ decays 
using the form-factor parametrizations discussed earlier.
Section~\ref{sec:FF0-T-fits} describes the calculation 
of the form factors $f_0 (q^2)$ and $f_T (q^2)$ 
for the $B \to \pi$ transition, using Lattice data 
as input in the high-$q^2$ region and the $z$-expansion 
to extrapolate it to low-$q^2$. 
Section~\ref{sec:B-pi-mu-mu-HQS-limit} contains the calculation 
of the dilepton invariant-mass spectrum in the low-$q^2$ region, 
using methods based on the heavy-quark symmetry. 
In Section~\ref{sec:entire-region}, we present 
the dilepton invariant-mass spectrum in the entire $q^2$-region 
as well as the partial decay rates, integrated 
over eight different $q^2$-intervals. 
A summary and outlook are given in Section~\ref{sec:summary}.

\section{\label{sec:B-pi-mu-mu-basis} 
The $B^+ \to \pi^+ \ell^+ \ell^-$ Decay
}

The effective weak Hamiltonian encompassing 
the transitions $b \to d \, \ell^+ \ell^-$  
($\ell^\pm = e^\pm$, $\mu^\pm$, or~$\tau^\pm)$,   
in the Standard Model (SM) can be written 
as follows~\cite{Buchalla:1995vs}:  
%
\begin{eqnarray}
{\cal H}^{b \to d}_{\rm eff} & = & 
\frac{4 G_F}{\sqrt 2} \left [ V_{ud} V_{ub}^* \left ( 
C_1 \, {\cal O}^{(u)}_1 + C_2 \, {\cal O}^{(u)}_2  
\right ) 
\right. 
\label{eq:Heff} \\ 
& + & 
\left. 
V_{cd} V_{cb}^* \left ( 
C_1 \, {\cal O}_1  + C_2 \, {\cal O}_2  
\right ) 
- V_{td} V_{tb}^* \sum\limits_{i=3}^{10} C_i \, {\cal O}_i  
\right ] ,
\nonumber 
\end{eqnarray}
where~$G_F$ is the Fermi constant, 
$V_{q_1 q_2}$ are the CKM matrix elements 
which satisfy the unitary condition
$V_{ud} V_{ub}^* + V_{cd} V_{cb}^* + V_{td} V_{tb}^* = 0$  
(it can be used to eliminate one combination). 
In contrast to the $b \to s$ transitions, all three 
terms in the unitarity relation are of the same order 
in~$\lambda$ ($V_{ub}^* V_{ud} \sim V_{cb}^* V_{cd} 
\sim V_{tb}^* V_{td} \sim \lambda^3$), with
$\lambda = \sin \theta_{12}\simeq 0.2232$~\cite{Beringer:1900zz}.  

The local operators appearing in Eq.~(\ref{eq:Heff})  
are the dimension-six operators, 
and are defined at an arbitrary scale~$\mu$ 
as follows~\cite{Chetyrkin:1996vx,Bobeth:1999mk}:  
\begin{subequations} 
\label{eq:operator-set} 
\begin{eqnarray}
&& 
{\cal O}^{(u)}_1 =  
\left ( \bar d_L \gamma_\mu T^A u_L \right ) 
\left ( \bar u_L \gamma^\mu T^A b_L \right ) , \hspace*{7mm}
\label{eq:O1u-operator} \\ 
&& 
{\cal O}^{(u)}_2  = 
\left ( \bar d_L \gamma_\mu u_L \right ) 
\left ( \bar u_L \gamma^\mu b_L \right ) , 
\label{eq:O2u-operator} \\   
&& 
{\cal O}_1 =  
\left ( \bar d_L \gamma_\mu T^A c_L \right ) 
\left ( \bar c_L \gamma^\mu T^A b_L \right ) ,  \hspace*{11mm}
\label{eq:O1-operator} \\ 
&&  
{\cal O}_2 = 
\left ( \bar d_L \gamma_\mu c_L \right ) 
\left ( \bar c_L \gamma^\mu b_L \right ) , 
\label{eq:O2-operator} \\   
&& 
{\cal O}_3 =  
\left ( \bar d_L \gamma_\mu b_L \right ) 
\mbox{$\sum_q$} \left ( \bar q \gamma^\mu q \right ) ,  \hspace*{20mm}
\label{eq:O3-operator} \\ 
&& 
{\cal O}_4 = 
\left ( \bar d_L \gamma_\mu T^A b_L \right ) 
\mbox{$\sum_q$} \left ( \bar q \gamma^\mu T^A q \right ) , 
\label{eq:O4-operator} \\ 
&& 
{\cal O}_5 =  
\left ( \bar d_L \gamma_\mu \gamma_\nu \gamma_\rho 
        b_L \right ) 
\mbox{$\sum_q$} \left ( \bar q \gamma^\mu \gamma^\nu 
               \gamma^\rho q \right ),  \hspace*{4mm} 
\label{eq:O5-operator} \\ 
&& 
{\cal O}_6 = 
\left ( \bar d_L \gamma_\mu \gamma_\nu \gamma_\rho 
        T^A b_L \right ) 
\mbox{$\sum_q$} \left ( \bar q \gamma^\mu \gamma^\nu 
               \gamma^\rho T^A q \right ) , 
\label{eq:O6-operator} \\
&& 
{\cal O}_7 = \frac{e \, m_b}{g_{\rm s}^2}   
\left ( \bar d_L \sigma^{\mu \nu} b_R \right ) F_{\mu \nu} ,  \hspace*{21mm}
\label{eq:O7-operator} \\ 
&& 
 {\cal O}_8 = \frac{m_b}{g_{\rm s}} 
\left ( \bar d_L \sigma^{\mu \nu} T^A b_R \right ) G_{\mu \nu}^A 
\label{eq:-O8-operator} \\ 
&& 
{\cal O}_9 = \frac{e^2}{g_{\rm s}^2}   
\left ( \bar d_L \gamma^\mu b_L \right ) 
\sum_\ell \left ( \bar \ell \gamma_\mu \ell \right ) ,  \hspace*{14mm}
\label{eq:O9-operator} \\ 
&& 
{\cal O}_{10}  =  \frac{e^2}{g_{\rm s}^2}
\left ( \bar d_L \gamma^\mu b_L \right ) 
\sum_\ell \left ( \bar \ell \gamma_\mu \gamma_5 \ell \right ),
\label{eq:O10-operator}
\end{eqnarray}
\end{subequations}
where~$e$ is the electric elementary charge, 
$g_{\rm s}$ is the strong coupling, 
$T^A$ ($A = 1, \ldots, N_c^2 - 1$) are the generators 
of the color $SU (N_c)$-group with $N_c = 3$, 
$\sigma_{\mu \nu} = 
i \left ( \gamma_\mu \gamma_\nu - \gamma_\nu \gamma_\mu \right )/2$, 
the subscripts~$L$ and~$R$ refer to the left- 
and right-handed components of the fermion fields, 
$\psi_{L,R} (x) = \left (1 \mp \gamma_5 \right) \psi (x)/2$,
$F_{\mu\nu}$ and $G^A_{\mu\nu}$ 
are the photon and gluon fields, respectively, 
and $m_b$ is the $b$-quark mass. 
(The terms in the~${\cal O}_7$ and~${\cal O}_8$ operators 
proportional to the $d$-quark mass~$m_d$ are omitted 
as their contributions to the amplitudes are 
suppressed by the ratio $m_d/m_b \sim 10^{-3}$
and negligible at the present level of accuracy). 
Sums over~$q$ and~$\ell$ denote sums over all  
quarks (except the $t$-quark) and charged 
leptons, respectively.

\begin{figure}[tb] 
\begin{center}
\includegraphics[width = 0.4\textwidth]{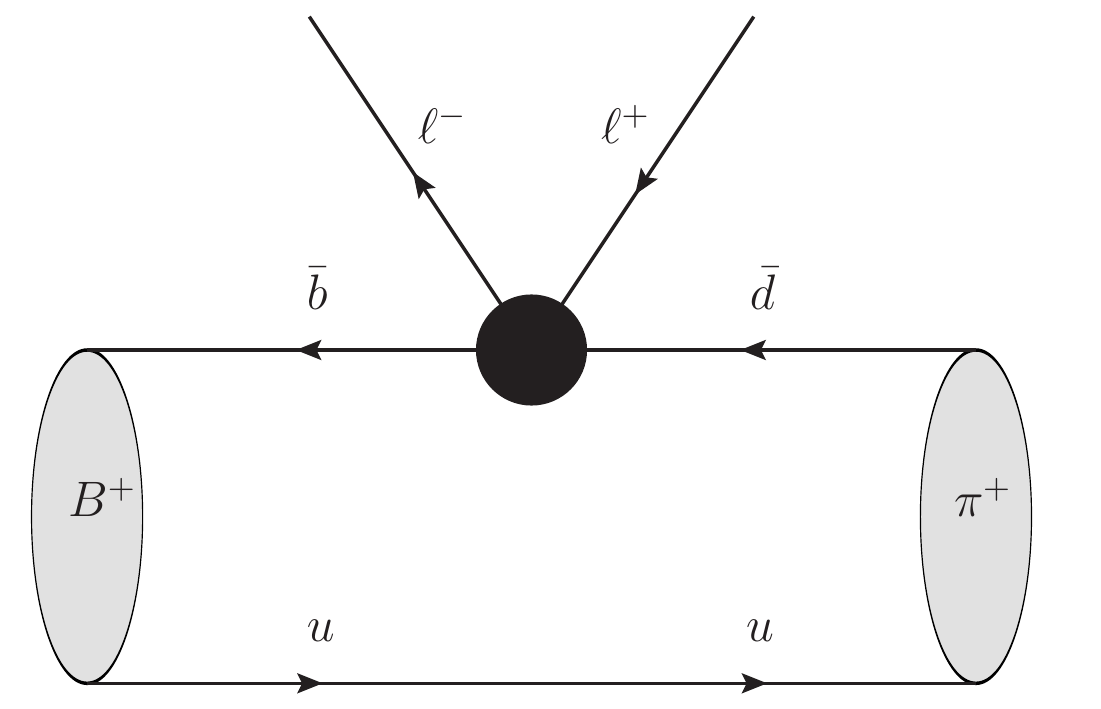}
\end{center}
\caption{\label{fig:B-pi-ell-ell}
Feynman diagram for the $B^+ \to \pi^+ \ell^+ \ell^-$ decay. 
} 
\end{figure}

The Wilson coefficients $C_i (\mu)$ ($i = 1, \ldots, 10$) 
depending on the renormalization scale~$\mu$ are 
calculated at the matching scale $\mu_W \sim M_W$, 
the $W$-boson mass, as a perturbative expansion in the 
strong coupling constant $\alpha_s (\mu_W)$~\cite{Bobeth:1999mk}:  
%
\begin{eqnarray} 
C_i (\mu_W) & = & C^{(0)}_i (\mu_W) + 
\frac{\alpha_s (\mu_W)}{4 \pi} \, C^{(1)}_i (\mu_W) 
\nonumber \\ 
& + & 
\left ( \frac{\alpha_s (\mu_W)}{4 \pi} \right )^2 
C^{(2)}_i (\mu_W) + \ldots , 
\label{eq:WC-matching}
\end{eqnarray} 
%
and can be evolved to a lower scale~$\mu_b \sim m_b$ 
using the anomalous dimensions of the above operators 
to the NNLL order~\cite{Bobeth:1999mk}: 
%
\begin{eqnarray} 
\gamma_i & = & 
  \frac{\alpha_s (\mu_W)}{4 \pi} \, \gamma^{(0)}_i + 
  \left ( \frac{\alpha_s (\mu_W)}{4 \pi} \right )^2 \gamma^{(1)}_i  
\nonumber \\ 
& + & 
  \left ( \frac{\alpha_s (\mu_W)}{4 \pi} \right )^3 \gamma^{(2)}_i 
+ \ldots .  
\label{eq:AD-expansion}
\end{eqnarray} 

Feynman diagram of the $B^+ \to \pi^+ \ell^+ \ell^-$ decay 
is displayed in Fig.~\ref{fig:B-pi-ell-ell} in which 
the solid blob represents the effective Hamiltonian 
${\cal H}^{b \to d}_{\rm eff}$~(\ref{eq:Heff}). 
The hadronic matrix elements of the operators~${\cal O}_i$ 
between the $B$- and $\pi$-meson states are expressed 
in terms of three independent form factors $f_+ (q^2)$, 
$f_0 (q^2)$ and $f_T (q^2)$ as follows~\cite{Beneke:2000wa}:
\begin{eqnarray}
\langle \pi (p_\pi) | \bar b \gamma^\mu d | B (p_B) \rangle 
& = & 
f_+ (q^2) \left [ p_B^\mu + p_\pi^\mu - 
\frac{m_B^2 - m_\pi^2}{q^2} \, q^\mu \right] 
\nonumber \\ 
& + & 
f_0 (q^2) \, 
\frac{m_B^2 - m_\pi^2}{q^2} \, q^\mu,  
\label{eq:ME-vector} 
\end{eqnarray}
\begin{eqnarray}
\langle \pi (p_\pi) | \bar b \sigma^{\mu\nu} q_\nu d | B (p_B) \rangle 
& = & 
\frac{i f_T (q^2)}{m_B + m_\pi} 
\label{eq:ME-tensor} \\ 
&& \hspace*{-10mm} 
\times  
\left[ q^2 \left ( p_B^\mu + p_\pi^\mu \right ) - 
       \left ( m_B^2 - m_\pi^2 \right ) q^\mu \right] , 
\nonumber 
\end{eqnarray}
where~$p_B^\mu$ and~$p_\pi^\mu$ are the four-momenta 
of the $B$- and $\pi$-mesons, respectively, 
$m_B$ and $m_\pi$ are their masses, and 
$q^\mu = p_B^\mu - p_\pi^\mu$ is the momentum 
transferred to the lepton pair.  
The $B \to \pi$ transition form factors 
$f_+ (q^2)$, $f_0 (q^2)$ and $f_T (q^2)$ 
are scalar functions whose shapes are determined 
by using non-perturbative methods. 
Of these, using the isospin symmetry, 
$f_+ (q^2)$ can also be obtained by performing 
a phenomenological analysis of the existing experimental 
data on the charged-current semileptonic decays 
$B \to \pi \ell \nu_\ell$. In the large-recoil 
(low-$q^2$) limit, these form factors are related 
by the heavy-quark symmetry, as discussed below.


The differential branching fraction in the dilepton 
invariant mass~$q^2$ can be expressed as follows:
%
\begin{eqnarray}
\frac{d {\cal B} \left ( B^+ \to \pi^+ \ell^+ \ell^- \right )}{d q^2} 
& = & 
\frac{G_F^2 \alpha_{\rm em}^2 \tau_B}{1024 \pi^5 m_B^3} \,  
|V_{tb} V_{td}^*|^2  
\label{eq:B-pi-ell-ell-LIMDF} \\
& \times & 
\sqrt{\lambda (q^2)} \sqrt{1 - \frac{4 m_\ell^2}{q^2}} \, F (q^2) , 
\nonumber 
\end{eqnarray}
%
where $\alpha_{\rm em}$ is the fine-structure constant, 
$m_\ell$ is the lepton mass, 
$\tau_B$ is the $B$-meson lifetime, 
\begin{equation}  
\lambda (q^2) = 
\left ( m_B^2 + m_\pi^2 - q^2 \right )^2 - 4 m_B^2 m_\pi^2 
\label{eq:lambda}
\end{equation} 
is the kinematic function encountered 
in three-body decays (the triangle function),  
and $F (q^2)$ is the dynamical function encoding 
the Wilson coefficients and the form factors:  
\begin{eqnarray}  
F (q^2) & = & \frac{2}{3} \, \lambda(q^2) 
\left ( 1 + \frac{2 m_\ell^2}{q^2} \right ) 
\label{eq:dynamical-function-def} \\ 
& \times & 
\left | C_9^{\rm eff} (q^2) \, f_+ (q^2) + 
\frac{2 m_b}{m_B + m_\pi} \, C_7^{\rm eff} (q^2) \, 
f_T (q^2) \right |^2 
\nonumber \\ 
& + & 
\frac{2}{3} \, \lambda(q^2) 
\left ( 1 - \frac{4 m_\ell^2}{q^2} \right ) 
\left | C_{10}^{\rm eff} \, f_+ (q^2) \right |^2 
\nonumber \\ 
& + & 
\frac{4 m_\ell^2}{q^2} 
\left ( m_B^2 - m_\pi^2 \right)^2 
\left | C_{10}^{\rm eff} \, f_0 (q^2) \right |^2 .  
\nonumber 
\end{eqnarray} 
Note that the last term 
in Eq.~(\ref{eq:dynamical-function-def})  
containing the form factor $f_0 (q^2)$ 
is strongly suppressed by the mass ratio $m_\ell^2/q^2$ 
for the electron- or muon-pair production over the 
most of the dilepton invariant-mass spectrum and
will not be needed in our numerical estimates. 
The dynamical function~(\ref{eq:dynamical-function-def})  
contains the effective Wilson coefficients 
$C_7^{\rm eff} (q^2)$, $C_9^{\rm eff} (q^2)$ 
and~$C_{10}^{\rm eff}$ which are specific 
combinations of the Wilson coefficients entering 
the effective Hamiltonian~(\ref{eq:Heff}).   
To the NNLO approximation, the effective Wilson 
coefficients are given by~\cite{Bobeth:1999mk,%
Asatrian:2001de,Asatryan:2001zw,Ali:2002jg,Asatrian:2003vq}:  
\begin{eqnarray}
C^{\rm eff}_7 (q^2) & = & A_7 - 
\frac{\alpha_s (\mu)}{4 \pi} 
\label{eq:C7-eff} \\
& \times &
\left [  
C_1^{(0)} F_1^{(7)} (s) + 
C_2^{(0)} F_2^{(7)} (s) + 
A_8^{(0)} F_8^{(7)} (s) 
\right ] 
\nonumber \\  
& + & \lambda_u \frac{\alpha_s (\mu)}{4 \pi} 
\left [ 
C_1^{(0)} \left ( F_{1, u}^{(7)} (s) - F_1^{(7)} (s) \right ) 
\right. 
\nonumber \\  
&& \hspace*{11mm}  
+ \left. 
C_2^{(0)} \left ( F_{2, u}^{(7)} (s) - F_2^{(7)} (s) \right ) 
\right ] , 
\nonumber  
\end{eqnarray}
\begin{eqnarray}
C^{\rm eff}_9 (q^2) & = &  
A_9 + T_9 \, h (m_c^2, q^2) 
\label{eq:C9-eff} \\
& + & 
U_9 \, h (m_b^2, q^2) + W_9 \, h (0, q^2) 
- \frac{\alpha_s (\mu)}{4 \pi} 
\nonumber \\ 
& \times & 
\left [  
C_1^{(0)} F_1^{(9)} (s) + 
C_2^{(0)} F_2^{(9)} (s) + 
A_8^{(0)} F_8^{(9)} (s) 
\right ] 
\nonumber \\ 
& + & \lambda_u 
\left [ \frac{4}{3} \, C_1 + C_2 \right ] 
\left [ h (m_c^2, q^2) - h (0, q^2) \right ]  
\nonumber \\
& + & \lambda_u \frac{\alpha_s (\mu)}{4 \pi} \left [ 
C_1^{(0)} \left ( F_{1, u}^{(9)} (s) - F_1^{(9)} (s) \right ) 
\right. 
\nonumber \\  
&& \hspace*{11mm}  
+ \left. 
C_2^{(0)} \left ( F_{2, u}^{(9)} (s) - F_2^{(9)} (s) \right ) 
\right ] , 
\nonumber 
\end{eqnarray}
\begin{equation}
C_{10}^{\rm eff} = \frac{4 \pi}{\alpha_s (\mu)} \, C_{10} ,  
\label{eq:C10-eff}
\end{equation}
where $s = q^2/m_B^2$ is the reduced momentum squared  
of the lepton pair. The quantity~$\lambda_u$ above is 
the ratio of the CKM matrix elements, defined as follows: 
\begin{equation}
\lambda_u \equiv \frac{V_{ub} V_{ud}^*}{V_{tb} V_{td}^*} = 
- \frac{R_b}{R_t} \, e^{i \alpha} , 
\label{eq:lambda-u-def}
\end{equation}
which is expressed in terms of the apex angle~$\alpha$ 
and the sides $R_t = \sqrt{( 1 - \bar\rho)^2 + \bar\eta^2}$ and 
$R_b = \sqrt{\bar\rho^2 + \bar\eta^2}$~\cite{Beringer:1900zz} 
of the CKM unitarity triangle, where~$\bar\rho$ and~$\bar\eta$ 
are the perturbatively improved Wolfenstein 
parameters~\cite{Wolfenstein:1983yz} of the CKM matrix.
The usual procedure is to include an additional term usually 
denoted by~$Y (q^2)$~\cite{Ali:1999mm,Ali:2002jg} into the 
$C_9^{\rm eff} (q^2)$ Wilson coefficient~(\ref{eq:C9-eff})   
which effectively accounts for the resonant states 
(mostly charmonia decaying into the lepton pair).    
The study of the long-distance effects based both 
on theoretical tools and experimental data on 
the two-body hadronic decays $B \to K^{(*)} + V$, 
where~$V$ is a vector meson decaying into the 
lepton pair $V \to \ell^+ \ell^-$, was undertaken 
recently in the context of the FCNC semileptonic 
decays $B \to K^{(*)} \ell^+ \ell^-$~\cite{%
Khodjamirian:2010vf,Khodjamirian:2012rm,Khodjamirian:2013iaa}. 
The resonant contributions can be largely removed 
by a stringent cut, but they may have a moderate 
impact also away from the resonant region and are 
included in the analysis of the data. 
Similar analysis can be undertaken for the 
$B \to \left ( \pi, \rho, \omega \right ) \ell^+ \ell^-$ decays 
also, but is not yet performed~\cite{Khodjamirian:2013iaa}.  
We concentrate here on the short-distance 
part of the differential branching ratio.

Following the prescription of Ref.~\cite{Ali:2002jg}, 
the terms~$\omega_i (s)$ accounting for the 
bremsstrahlung corrections necessary for the inclusive 
$B \to (X_s, \, X_d) \, \ell^+ \ell^-$ decays are omitted 
and, the following set of auxiliary functions is used: 
\begin{eqnarray}
A_7 (\mu) & = & \frac{4 \pi}{\alpha_s (\mu)} \, C_7 (\mu) - 
\frac{1}{3} \, C_3 (\mu) - \frac{4}{9} \, C_4 (\mu) 
\label{eq:A7} \\
& - &  
\frac{20}{3} \, C_5 (\mu) - \frac{80}{9} \, C_6 (\mu) , 
\nonumber 
\end{eqnarray}
\begin{eqnarray}
A_8 (\mu) & = & \frac{4 \pi}{\alpha_s (\mu)} \, C_8 (\mu) +  
C_3 (\mu) - \frac{1}{6} \, C_4 (\mu) 
\label{eq:A8} \\
& + & 
20 \, C_5 (\mu) - \frac{10}{3} \, C_6 (\mu) , 
\nonumber 
\end{eqnarray}
\begin{eqnarray}
A_9 (\mu) & = & \frac{4 \pi}{\alpha_s (\mu)} \, C_9 (\mu) +  
\sum_{i = 1}^6 C_i (\mu) \, \gamma_{i 9}^{(0)} \, \ln\frac{m_b}{\mu}
\label{eq:A9} \\
& + & 
\frac{4}{3} \, C_3 (\mu) + \frac{64}{9} \, C_5 (\mu) + 
\frac{64}{27} \, C_6 (\mu) , 
\nonumber 
\end{eqnarray}
\begin{equation}
T_9 (\mu) = \frac{4}{3} \, C_1 (\mu) + C_2 (\mu) + 
6 \, C_3 (\mu) + 60 \, C_5 (\mu) ,  
\label{eq:T9} 
\end{equation}
%
\begin{equation} 
U_9 (\mu) = 
- \frac{7}{2} \, C_3 (\mu) - \frac{2}{3} \, C_4 (\mu) 
- 38 \, C_5 (\mu) - \frac{32}{3} \, C_6 (\mu) , 
\label{eq:U9} 
\end{equation} 
\begin{equation} 
W_9 (\mu) = 
- \frac{1}{2} \, C_3 (\mu) - \frac{2}{3} \, C_4 (\mu) 
- 8 \, C_5 (\mu) - \frac{32}{3} \, C_6 (\mu) ,   
\label{eq:W9} 
\end{equation} 
%
where the required elements of the anomalous 
dimension matrix~$\gamma^{(0)}_{ij}$ can be 
read off from Ref.~\cite{Bobeth:1999mk}.  
The numerical values of the scale-dependent 
functions specified above at three representative 
scales $\mu = 2.45$~GeV, $\mu = 4.90$~GeV 
and $\mu = 9.80$~GeV are presented 
in Table~\ref{tab:Wilson-coeff}. 
\begingroup 
\squeezetable 
\begin{table}[tb] 
\caption{\label{tab:Wilson-coeff}
Wilson coefficients~$C_1$, $C_2$, $C^{\rm eff}_{10}$,   
and the combinations of the Wilson coefficients 
specified in Eqs.~(\ref{eq:A7})--(\ref{eq:W9}), are shown 
at three representative renormalization 
scales: $\mu_b = 2.45$~GeV, $\mu_b = 4.90$~GeV 
and $\mu_b = 9.80$~GeV. 
The strong coupling $\alpha_s (\mu)$ 
is evaluated by the three-loop expression 
in the $\overline{\rm MS}$ scheme 
with five active flavors and 
$\alpha_s (M_Z) = 0.1184$~\cite{Beringer:1900zz}.   
The entries correspond to the top-quark mass 
$m_t = 175$~GeV.  
The superscript~$(0)$ denotes the lowest 
order contribution while a quantity with 
the superscript~$(1)$ is a perturbative 
correction of order~$\alpha_s$, and  
$X = X^{(0)} + X^{(1)}$.  
}
\begin{center}
\begin{tabular}{cccc}
\hline\hline 
& $\mu = 2.45$~GeV & $\mu = 4.90$~GeV & $\mu = 9.80$~GeV  
\\ \hline 
$\alpha_s (\mu)$ & 0.269 & 0.215 & 0.180 \\ 
$(C^{(0)}_1$, $C^{(1)}_1$) & 
$(-0.707, \, 0.241)$ & $(-0.492, \, 0.207)$ & $(- 0.330, \, 0.184)$ \\ 
$(C^{(0)}_2$, $C^{(1)}_2$) & 
$(1.047, \, -0.028)$ & $(1.024, \, -0.017)$ & $(1.011, \, 0.010)$ \\ 
$(A^{(0)}_7$, $A^{(1)}_7$) & 
$(-0.355, \, 0.025)$ & $(-0.313, \, 0.010)$ & $(- 0.278, \, -0.001)$ \\ 
$A^{(0)}_8$ & 
$-0.164$ & $-0.148$ & $-0.134$ \\ 
$(A^{(0)}_9$, $A^{(1)}_9$) & 
$(4.299, \, -0.237)$ & $(4.171, \, -0.053)$ & $(4.164, \, 0.090)$ \\ 
$(T^{(0)}_9$, $T^{(1)}_9$) & 
$(0.101, \, 0.280)$ & $(0.367, \, 0.251)$ & $(0.571, \, 0.231)$ \\ 
$(U^{(0)}_9$, $U^{(1)}_9$) & 
$(0.046, \, 0.023)$ & $(0.033, \, 0.015)$ & $(0.023, \, 0.010)$ \\ 
$(W^{(0)}_9$, $W^{(1)}_9$) & 
$(0.045, \, 0.016)$ & $(0.032, \, 0.012)$ & $(0.022, \, 0.008)$ \\ 
$(C^{{\rm eff} (0)}_{10}$, $C^{{\rm eff} (1)}_{10}$) & 
$(-4.560, \, 0.378)$ & $(-4.560, \, 0.378)$ & $(-4.560, \, 0.378)$ \\ 
\hline\hline 
\end{tabular}
\end{center}
\end{table} 
\endgroup 
In Eq.~(\ref{eq:C9-eff}), $m_c$ and $m_b$ are 
the $c$- and $b$-quark masses, respectively, 
the masses of the light $u$-, $d$-, and $s$-quarks 
are neglected, and the standard one-loop function~$h (z, s)$ 
is used~\cite{Buchalla:1995vs} ($x = 4 z /s$): 
\begin{eqnarray}
h (z, s) & = & - \frac{4}{9} \ln \frac{z}{\mu^2} 
+ \frac{8}{27} + \frac{4}{9} \, x -
\frac{2}{9} \left ( 2 + x \right ) \sqrt{|1 - x|} 
\nonumber \\
& \times &  
\left\{
\begin{array}{ll}
\ln \left | \displaystyle
\frac{1 + \sqrt{1 - x}}{1 - \sqrt{1 - x}} 
\right | - i \pi, & \mbox{ for} \,  x < 1, \\
2 \arctan (1/\sqrt{x - 1}), &  \mbox{ for}  \,  x \ge 1.  
\end{array} 
\right. 
\label{h-function} 
\end{eqnarray}

\begin{figure*}[tb] 
\begin{center}
\includegraphics[width = 0.45\textwidth]{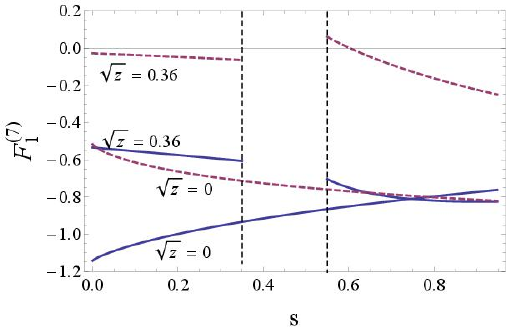} 
\hfill 
\includegraphics[width = 0.45\textwidth]{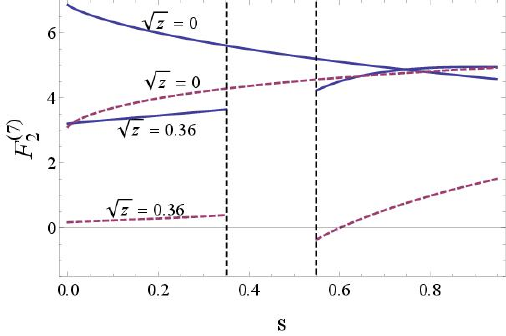} 
\\[3mm]
\includegraphics[width = 0.45\textwidth]{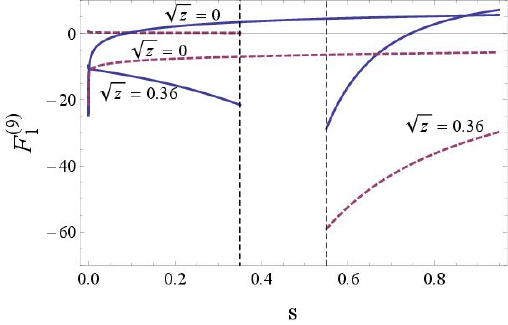}  
\hfill 
\includegraphics[width = 0.45\textwidth]{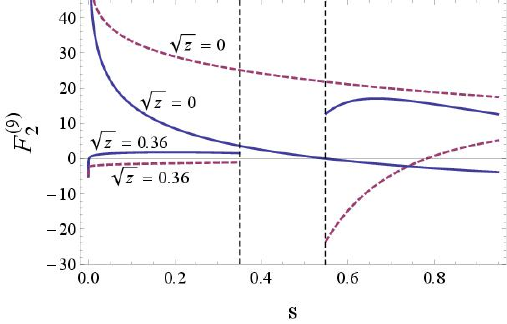} 
\end{center} 
\caption{\label{fig:F-12-79-corrections}  
(Color online.) 
The real (solid lines) and imaginary 
(dotted lines) parts of the functions 
$F^{(7)}_{1,2} (s)$ (top two frames) 
and $F^{(9)}_{1,2} (s)$ (bottom two frames)   
at the scale $\mu = m_b$.  
For plotting the curves with $\sqrt z = 0$, 
the exact analytic expressions~\cite{Seidel:2004jh} 
were used. For non-zero values of~$\sqrt z$,  
the analytic two-loop expressions 
obtained as double expansions in~$\sqrt z$ 
and~$s$~\cite{Asatrian:2001de,Asatryan:2001zw} 
are used in plotting these functions in the region $s \leq 0.35$, 
whereas the expansions in~$\sqrt z$ and~$1 - s$~\cite{Greub:2008cy} 
are used in the range $0.55 < s < 1$. 
For these curves, we have fixed $\sqrt z = 0.36$. 
}
\end{figure*}

The renormalized $\alpha_s$-corrections 
$F^{(7)}_{1,2} (s)$ and $F^{(9)}_{1,2} (s)$ 
to the $b \to s \, \ell^+ \ell^-$ matrix element 
originated by the ${\cal O}_1$- and 
${\cal O}_2$-operators from the effective 
Hamiltonian~(\ref{eq:Heff}) are known analytically both 
in the small-$q^2$~\cite{Asatrian:2001de,Asatryan:2001zw} 
and large-$q^2$~\cite{Greub:2008cy} domains 
of the lepton invariant mass squared 
as expansions in $\sqrt z = m_c/m_b$.
Note that to obtain the invariant-mass spectrum 
and forward-backward asymmetry in the inclusive 
$B \to X_s \ell^+ \ell^-$ decays the 
$F_{1,2,8}^{(7)} (s)$ and~$F_{1,2,8}^{(9)} (s)$ 
functions were expressed in terms of master integrals 
and evaluated numerically~\cite{Ghinculov:2003qd}. 
The functions $F_{1 (2), u}^{(7)} (s)$ and 
$F_{1 (2),u}^{(9)} (s)$ which are important 
in the $b \to d \, \ell^+ \ell^-$ transitions 
were also calculated analytically first as an 
expansion in powers of~$s$~\cite{Asatrian:2003vq} 
and later exactly~\cite{Seidel:2004jh} from which 
the later expressions are used by us as we are 
considering the $B \to \pi \ell^+ \ell^-$ decay 
in the entire $q^2$-region.

The functions $F^{(7)}_{1,2} (s)$ (the top two frames)
and $F^{(9)}_{1,2} (s)$ (the bottom two frames) 
are presented in Fig.~\ref{fig:F-12-79-corrections}
at the scale $\mu = m_b$ and $\sqrt z = 0.36$. 
The real and imaginary parts of these functions 
are shown by the solid and dashed lines, respectively.   
The functions $F^{(7)}_{1,2} (s)$ and $F^{(9)}_{1,2} (s)$ 
at $\sqrt z = 0$, which are obtained analytically 
in Ref.~\cite{Seidel:2004jh}, are also shown 
 in Fig.~\ref{fig:F-12-79-corrections}. 
The vertical dashed lines specify the $s$-region  
where the expansions no longer hold.      
As the correct analytical functions in this region 
are not known for realistic value of~$\sqrt z$,  
we have extrapolated the known analytic expressions 
from above and below (i.\,e., using expansions in~$s$ 
and~$1 - s$) to a point in the intermediate region 
where the differential branching fraction has 
a minimal discontinuity. This allow us to get 
an approximate estimate of the perturbative part 
of the differential branching fraction in the gap 
between the $J/\psi$- and $\psi (2S)$-resonances. 

In the analysis we also used the renormalized 
$\alpha_s$-corrections $F^{(7,9)}_8 (s)$ 
from the ${\cal O}_8$-operator valid 
in the full kinematic $q^2$-domain 
($0 \le s \le 1$)~\cite{Greub:2008cy}: 
\begin{eqnarray}  
F^{(7)}_8 (s) & = &  
\frac{4 \pi^2}{27} \, \frac{2 + s}{(1 - s)^4} - 
\frac{16 (2 + s)}{3 (1 - s)^4} \arcsin^2 \frac{\sqrt s}{2} 
\label{eq:F-78} \\ 
& - & 
\frac{8 \sqrt{s (4 - s)}}{9 (1 - s)^3}  
\left ( 9 - 5 s + 2 s^2 \right ) \arcsin \frac{\sqrt s}{2} 
\nonumber \\ 
& - & 
\frac{4 (11 - 16 s + 8 s^2)}{9 (1 - s)^2} - 
\frac{8 s \ln s}{9 (1 - s)} 
\nonumber \\
& - & 
\frac{8 i \pi}{9} - 
\frac{32}{9} \ln \frac{\mu}{m_b} , 
\nonumber 
\end{eqnarray}
\begin{eqnarray}  
F^{(9)}_8 (s) & = &  
- \frac{8 \pi^2}{27} \, \frac{4 - s}{(1 - s)^4} 
+ \frac{8 (5 - 2 s)}{9 (1 - s)^2} 
\label{eq:F-98} \\  
& + & 
\frac{16 \sqrt{4 - s}}{9 \sqrt s \, (1 - s)^3}  
\left ( 4 + 3 s - s^2 \right ) \arcsin \frac{\sqrt s}{2} 
\nonumber \\ 
& + & 
\frac{32 (4 - s)}{3 (1 - s)^4} \arcsin^2 \frac{\sqrt s}{2} + 
\frac{16 \ln s}{9 (1 - s)} ,  
\nonumber 
\end{eqnarray}
where the $b$-quark mass~$m_b$ is assumed to be the pole mass. 

To perform the numerical analysis one needs to know the 
$B \to \pi$ transition form factors $f_+ (q^2)$, 
$f_0 (q^2)$ and $f_T (q^2)$ in the entire kinematic range: 
\begin{equation} 
4 m_\ell^2 \le q^2 \le \left ( m_B - m_\pi \right )^2 .  
\label{eq:q2-bounds}
\end{equation}
Their model-independent determination is the main aim of this 
paper, which is described in detail in subsequent sections. 



\section{\label{sec:FF-parametrizations}
Form-Factor Parametrizations
}

Several parametrizations of the 
$B \to \pi$ transition 
form factors~$f_+ (q^2)$, $f_0 (q^2)$ and~$f_T (q^2)$ 
have been proposed in the literature.
The four parametrizations of~$f_+ (q^2)$ discussed 
below have been used in the analysis of the semileptonic 
data on $B \to \pi \ell \nu_\ell$. All of them include 
at least one pole term at $q^2 = m_{B^*}^2$, 
where $m_{B^*} = 5.325$~GeV~\cite{Beringer:1900zz}   
is the vector $B^*$-meson mass. As far as this mass  
satisfies the condition $m_{B^*} < m_B + m_\pi$, 
i.\,e., it lies below the so-called continuum threshold, 
it should be included into the form factor as 
a separate pole. Other mesons and multi-particle states 
with the appropriate $J^P = 1^-$ quantum number 
can be described either by one or several 
poles or by some other rapidly convergent function,   
both effectively counting the continuum. 
The tensor form factor~$f_T (q^2)$ shows 
a similar qualitative behavior and its model 
function obeys the same shape as the vector one. 
The case of the scalar form factor~$f_0 (q^2)$ 
is different, as the first orbitally-excited 
scalar $B^{**}$-meson with $J^P = 0^+$ 
\footnote{it is expected to be somewhere within the signal 
called as the $B_J^* (5732)$ resonance~\cite{Beringer:1900zz} 
with the mass $m_{B_J^* (5732)} = (5698 \pm 8)$~MeV 
and width $\Gamma_{B_J^* (5732)} = (128 \pm 18)$~MeV 
which can be interpreted as stemming from several 
narrow and broad resonances. 
Approximately the same mass difference 
$m_{B_s^{**}} - m_{B_s} = (385 \pm 16 \pm 5 \pm 25)$~MeV 
in the $B_s$-meson sector was obtained by the 
HPQCD Collaboration~\cite{Gregory:2010gm}.} 
has the mass squared above the continuum threshold 
$t_0 = (m_B + m_\pi)^2 = 29.36$~GeV$^2$  
and, hence, it belongs to the continuum 
which makes~$f_0 (q^2)$  
regular at $q^2 = m_{B^*}^2$, in contrast 
to~$f_+ (q^2)$ and~$f_T (q^2)$.

\subsection{\label{ssec:FF-parametrization-BK}
The Becirevic-Kaidalov Parametrization
}

The form factor $f_+ (q^2)$ in the Becirevic-Kaidalov~(BK) 
parametrization~\cite{Becirevic:1999kt} can be written as follows: 
\begin{equation}
f_+ (q^2) = \frac{f_+ (0)}{
\left ( 1 - \hat q_*^2 \right )
\left ( 1 - \alpha_{\rm BK} \, \hat q_*^2 \right ) 
},
\label{eq:FFp-BK}
\end{equation}
where $\hat q_*^2 = q^2/m_{B^*}^2$. 
The fitted parameters are the form-factor normalization, 
$f_+ (0)$, and~$\alpha_{\rm BK}$ which defines 
the $f_+ (q^2)$ shape~\cite{Becirevic:1999kt}.
This parametrization is one of the simplest ones.
The shape of the tensor form factor $f_T (q^2)$ 
is the same~(\ref{eq:FFp-BK}) as it also has the pole 
at $q^2 = m_{B^*}^2$ below the continuum threshold.    
The scalar form factor $f_0 (q^2)$ was also introduced 
in its simplest form~\cite{Becirevic:1999kt}:  
\begin{equation}
f_0 (q^2) = \frac{f_+ (0)}{
1 - \hat q_*^2 / \beta_{\rm BK}  
},
\label{eq:FF0-BK}
\end{equation}
with the same normalization factor $f_+ (0)$ 
but a different effective pole position 
determined by the free parameter~$\beta_{\rm BK}$.

This form-factor parametrizations should be taken 
with caution, since the simple two-parameter shape 
is overly restrictive and has been argued to be 
inconsistent with the requirements from the 
Soft-Collinear Effective Theory (SCET)~\cite{Becher:2005bg}.

\subsection{\label{ssec:FF-parametrization-BZ}
The Ball-Zwicky Parametrization
}

The Ball-Zwicky (BZ) parametrization for the vector 
form factor $f_+ (q^2)$ is a modified form 
of the BK parametrization, given as~\cite{Ball:2004ye}:
%
\begin{eqnarray}
f_+ (q^2) & = & 
\frac{f_+ (0)}{1 - \hat q_*^2} \left [ 1 + 
\frac{r_{\rm BZ} \, \hat q_*^2}{
      1 - \alpha_{\rm BZ} \, \hat q_*^2} 
\right ] 
\label{eq:FFp-BZ} \\ 
& = & 
\frac{f_+ (0) \left [ 
      1 - \left ( \alpha_{\rm BZ} - r_{\rm BZ} \right ) \hat q_*^2 
              \right ]}
     {\left ( 1 - \hat q_*^2 \right ) 
      \left ( 1 - \alpha_{\rm BZ} \hat q_*^2 \right )} , 
\nonumber 
\end{eqnarray}
%
where the fitted parameters are $f_+ (0)$, $\alpha_{\rm BZ}$, 
and $r_{\rm BZ}$. $f_+ (0)$ sets again the normalization 
of the form factor, while~$\alpha_{\rm BZ}$ and~$r_{\rm BZ}$ 
define the shape~\cite{Ball:2004ye}.
In particular, for $\alpha_{\rm BZ} = r_{\rm BZ}$ 
one reproduces the BK parametrization~(\ref{eq:FFp-BK}). 
The same redefinition is also applied to the tensor form 
factor $f_T (q^2)$. In a similar way the scalar form 
factor $f_0 (q^2)$~(\ref{eq:FF0-BK}) can be modified by 
introducing its own second free parameter~$r^{(0)}_{\rm BZ}$.

\subsection{\label{ssec:FF-parametrization-BGL}
The Boyd-Grinstein-Lebed Parametrization
}

This parametrization was introduced for the form factors 
entering both the heavy-to-light~\cite{Boyd:1994tt} 
and heavy-to-heavy~\cite{Boyd:1995cf} transition   
matrix elements and used in the analysis of the 
semileptonic $B \to D^{(*)} \ell \nu_\ell$%
~\cite{Boyd:1995cf,Boyd:1995sq,Boyd:1997kz} and
$B \to \pi \ell \nu_\ell$~\cite{Boyd:1994tt,Boyd:1997qw} 
decays. The basic idea is to find an appropriate 
function~$z (q^2, q_0^2)$ in term of which the 
form factor can be written as a Taylor series with 
good convergence for all physical values of~$q^2$ 
so that the form factor can be well described 
by the first few terms in the expansion.  
The generalization of this parametrization 
to additional form factors entering rare 
semileptonic $B \to h_L \, \ell^+ \ell^-$, 
where~$h_L$ is the pseudoscalar~$K$- 
or the vector $\rho$- or $K^*$-mesons, 
and $B_s \to \phi \, \ell^+ \ell^-$ decays, 
was undertaken in~\cite{Bharucha:2010im}. 
As this will be our default parametrization 
in our analysis, we discuss it at some length.  

The following shape for the form factors  
$f_i (q^2)$ with $i = +, 0, T$ is suggested 
in the BGL parametrization~\cite{Boyd:1994tt}:   
\begin{equation}
f_i (q^2) = \frac{1}{P (q^2) \phi_i (q^2, q_0^2)} 
\sum\limits_{k = 0}^{k_{\rm max}}
a_k (q_0^2) \left [ z ( q^2, q_0^2) \right ]^k,
\label{eq:FF-BGL}
\end{equation}
where the following form for the function $z( q^2, q_0^2)$ is used:    
\begin{equation}
z (q^2, q_0^2) = 
\frac{\sqrt{m_+^2 - q^2} - \sqrt{m_+^2 - q_0^2}}
     {\sqrt{m_+^2 - q^2} + \sqrt{m_+^2 - q^2_0}} ,  
\label{eq:zqq0}
\end{equation}
with the pair-production threshold $m_+^2 = (m_B + m_\pi)^2$ 
and a free parameter~$q_0^2$. The function~$z (q^2, q_0^2)$
maps the entire range of~$q^2$ onto the unit disc  
$|z| \le 1$ in a way that the minimal physical value 
$z_{\rm min} = z (m_-^2, q_0^2)$ corresponds 
to the lowest hadronic recoil 
$q_{\rm max}^2 = m_-^2 = (m_B - m_\pi)^2$,   
the maximal value~$z_{\rm max}$ is reached at $q^2 = 0$, 
and $z (q^2, q_0^2)$ vanishes at $q^2 = q_0^2$. 
In early studies of the form factors, the parameter~$q_0^2$ was 
often taken to be $q_0^2 = m_-^2$~\cite{Boyd:1994tt,Boyd:1995cf}, 
so that $z_{\rm min} = 0$. In this case, the maximal value 
$z_{\rm max} = 0.52$ for the $B \to \pi \ell \nu_\ell$ 
decay is not small but enough to constrain the form 
factor~$f_+ (q^2)$~\cite{Lellouch:1995yv,Boyd:1997qw}. 
To decrease the value of~$z_{\rm max}$, and improve the 
convergence of the Taylor series in Eq.~(\ref{eq:FF-BGL}), 
it was proposed to take a smaller (optimal) value 
of~$q_0^2$ somewhere in the interval 
$0 < q_0^2 < m_-^2$~\cite{Boyd:1995tg}.    
In our analysis we make the choice $q^2_0 = 0.65 m_-^2$ 
following~\cite{delAmoSanchez:2010af}, 
so that $-0.34 < z (q^2, q_0^2) < 0.22$ 
in the entire range $0 < q^2 < m_-^2$.  

The proposed shape~(\ref{eq:FF-BGL}) for the form 
factor contains the so-called Blaschke factor~$P (q^2)$ 
which accounts for the hadronic resonances in the 
sub-threshold region $q^2 < m_+^2$. For the semileptonic 
$B \to \pi \ell \nu_\ell$ decay, where~$\ell$ is an 
electron or a muon, there is only the $B^*$-meson with 
the mass $m_{B^*} = 5.325$~GeV satisfying the 
sub-threshold condition and producing the pole 
in the form factor at $q^2 = m_{B^*}^2$. 
In this case, the Blaschke factor is simply 
$P (q^2) = z (q^2, m_{B^*}^2)$ for~$f_{+,T} (q^2)$, 
and $P (q^2) = 1$ for~$f_0 (q^2)$.  

The coefficients~$a_k$ ($k = 0, 1, \ldots, k_{\rm max}$) 
entering the Taylor series in Eq.~(\ref{eq:FF-BGL})  
are the parameters, which are determined by the fits of the data. 
The outer function $\phi_i (q^2, q_0^2)$ is an arbitrary 
analytic function, whose choice only affects particular values 
of the coefficients~$a_k$ and allows one to get a simple 
constraint from the dispersive bound~\cite{Boyd:1997qw}~\footnote{
The definition for~$f_0 (q^2)$ in accordance with 
Ref.~\cite{Bharucha:2010im} results in even stronger 
bound~$\sum_{k=0}^\infty a_k^2 \leq 1/3$.}: 
\begin{equation} 
\sum_{k=0}^\infty a_k^2 \le 1 . 
\label{eq:ak-restriction}  
\end{equation}
This restriction can be achieved with the following 
outer function~\cite{Arnesen:2005ez}: 
%
\begin{eqnarray}
\phi_i (q^2, q_0^2) & = & \sqrt{\frac{n_I}{K_i \chi_{f_i}^{(0)}}} \, 
\frac{(m_+^2 - q^2)^{(\alpha_i + 1)/4}}{(m_+^2 - q_0^2)^{1/4}}  
\label{eq:phiqq0} \\
& \times & 
\left ( \sqrt{m_+^2 - q^2} + \sqrt{m_+^2 - q_0^2} \right )
\nonumber \\  
& \times & 
\left ( \sqrt{m_+^2 - q^2} + \sqrt{m_+^2 - m_-^2} \right )^{\alpha_i/2}
\nonumber \\  
& \times & 
\left ( \sqrt{m_+^2 - q^2} + m_+ \right )^{-(3 + \beta_i)} ,  
\nonumber 
\end{eqnarray}
where $n_I = 3/2$ is the isospin factor, while the values of~$K_i$, 
$\alpha_i$ and~$\beta_i$ are collected in Table~\ref{tab:FF-numbers}. 
\begin{table}[tb] 
\caption{\label{tab:FF-numbers} 
Parameters entering the outer functions~$\phi_i (q^2, q_0^2)$ 
defined in Eq.~(\ref{eq:phiqq0}) with $i = +, 0, T$ 
in the $B \to \pi$ transition form factors. 
} 
\begin{center}
\begin{tabular}{lcccl} 
\hline  
$f_i$ & $K_i$ & $\alpha_i$ & $\beta_i$ & $\chi_i^{(0)} $
\\ \hline 
$f_+$ & $48 \pi$ & 3 & 2 & $7.005 \times 10^{-4}$~GeV$^{-2}$ \\  
$f_0$ & $16 \pi/(m_+^2 m_-^2)$ & 1 & 1 & $1.452 \times 10^{-2}$ \\  
$f_T$ & $48 \pi m_+^2$ & 3 & 1 & $1.811 \times 10^{-3}$~GeV$^{-2}$  
\\ \hline
\end{tabular}
\end{center}
\end{table} 
The numerical quantities~$\chi_{f_i}^{(0)}$ are obtained 
from the derivatives of the scalar functions entering 
the corresponding correlators calculated by the operator 
product expansion method~\cite{Boyd:1995tg,Boyd:1997qw,Bharucha:2010im}.  
In the two-loop order at the scale~$\mu_b$ 
they are as follows~\cite{Bharucha:2010im}: 
%
\begin{eqnarray} 
\chi_{f_+}^{(0)} & = & 
\frac{3}{32 \pi^2 m_b^2} \left ( 
1 + \frac{C_F \alpha_s (\mu_b)}{4 \pi} \, \frac{25 + 4 \pi^2}{6} 
\right ) 
\label{eq:chi-fp} \\  
& - & 
\frac{\langle \bar q q \rangle}{m_b^5} -  
\frac{\langle \alpha_s G^2 \rangle}{12 \pi m_b^6} +  
\frac{3 \langle \bar q G q \rangle}{m_b^7} ,   
\nonumber 
\end{eqnarray}
\begin{eqnarray} 
\chi_{f_0}^{(0)} & = & 
\frac{1}{8 \pi^2} \left ( 
1 + \frac{C_F \alpha_s (\mu_b)}{4 \pi} \, \frac{3 + 4 \pi^2}{6} 
\right ) 
\label{eq:chi-f0} \\  
& + & 
\frac{\langle \bar q q \rangle}{m_b^3} +  
\frac{\langle \alpha_s G^2 \rangle}{12 \pi m_b^4} -  
\frac{3 \langle \bar q G q \rangle}{2 m_b^5} ,   
\nonumber 
\end{eqnarray}
\begin{eqnarray} 
\chi_{f_T}^{(0)} & = & 
\frac{1}{4 \pi^2 m_b^2} \left ( 
1 + \frac{C_F \alpha_s (\mu_b)}{4 \pi} \left [  
\frac{10 + 2 \pi^2}{3} + 8 \ln \frac{m_b}{\mu_b}  
\right ] \right ) 
\nonumber \\  
& - & 
\frac{\langle \bar q q \rangle}{m_b^5} -  
\frac{\langle \alpha_s G^2 \rangle}{24 \pi m_b^6} +  
\frac{7 \langle \bar q G q \rangle}{2 m_b^7} ,   
\label{eq:chi-fT} 
\end{eqnarray}
where $C_F = 4/3$, and~$m_b$ 
is the mass of the $b$-quark in the loops which 
is identified with the $\overline{\rm MS}$ $b$-quark mass 
$\bar m_b (\bar m_b) = 4.18$~GeV~\cite{Beringer:1900zz}. 
For the evaluation of~$\chi_{f_i}^{(0)}$ it is enough to use 
the central values of the input parameters to get the overall 
numerical normalization factor for the form factors and the 
existing uncertainties in~$\chi_{f_i}^{(0)}$ are of not much 
consequence. The following input values are used: 
$\alpha_s (M_Z) = 0.1184 \pm 0.0007$~\cite{Beringer:1900zz}, 
$\langle \bar q q \rangle (1~{\rm GeV}) = 
- (1.65 \pm 0.15) \times 10^{-2}$~GeV$^3$,  
$\langle \bar q G q \rangle = 
\langle \bar q g_s \sigma^{\mu\nu} G_{\mu\nu}^A T^A q \rangle = 
m_0^2 \langle \bar q q \rangle$, 
$m_0^2  (1~{\rm GeV}) = (0.8 \pm 0.2)$~GeV$^2$, and  
$\langle (\alpha_s/\pi) \, G^2 \rangle = (0.005 \pm 0.004)$~GeV$^4$
from Ref.~\cite{Ioffe:2005ym}. 
While the mixed quark-gluon $\langle \bar q G q \rangle$ 
and the two-gluon $\langle (\alpha_s/\pi) \, G^2 \rangle$ condensates 
are practically scale-independent quantities~\cite{Ioffe:2005ym}, 
the strong coupling and the quark condensate have to be evolved 
to the scale of the $b$-quark mass where they have the values    
$\alpha_s (\bar m_b) = 0.227$ to the two-loop accuracy 
and $\langle \bar q q \rangle (\bar m_b) = - 0.023$~GeV$^3$.   
Numerical values of~$\chi_{f_i}^{(0)}$ are presented 
in Table~\ref{tab:FF-numbers}. They agree well (up to~5\%) 
with the ones presented in Table~2 of~\cite{Bharucha:2010im},
despite differences in the input parameters. 
Note that the BaBar Collaboration~\cite{delAmoSanchez:2010af} 
used approximately the same value 
$\chi_{f_+}^{(0)} = 6.889 \times 10^{-4}$~GeV$^{-2}$ 
in the analysis of the $B^0 \to \pi^+ \ell^- \nu_\ell$ decays.  
  
Having relatively small values of~$z (q^2, q_0^2)$ 
in the physical region of~$q^2$, the shape of the 
form factor can be well approximated by the truncated 
series at $k_{\rm max} = 2$ or~$3$~\cite{Boyd:1995sq}.

\subsection{\label{ssec:FF-parametrization-BCL}
The Bourrely-Caprini-Lellouch Parametrization
}

The problems with the from-factor asymptotic behavior 
at $|q^2| \to \infty$ and truncation of the Taylor series 
found in the BGL-parametrization~\cite{Becher:2005bg,Bourrely:2008za}  
were solved by the introduction of another representation 
of the series expansion (called the Simplified Series 
Expansion~--- SSE~\cite{Bharucha:2010im}).  
The shape suggested for the vector $f_+ (q^2)$ form 
factor~\cite{Bourrely:2008za} was extended to the other
two, scalar $f_0 (q^2)$ and tensor $f_T (q^2)$, form 
factors~\cite{Bharucha:2010im}: 
\begin{eqnarray}
f_+ (q^2) & = & \frac{1}{1 - \hat q_*^2} 
\sum\limits_{k = 0}^{k_{\rm max}} b^{(+)}_k (q_0^2)
\left [ z (q^2, q_0^2) \right ]^k , 
\label{eq:FFp-BCL} \\
f_0 (q^2) & = & \frac{m_B^2}{m_B^2 - m_\pi^2} 
\sum\limits_{k = 0}^{k_{\rm max}} b^{(0)}_k (q_0^2)
\left [ z (q^2, q_0^2) \right ]^k , 
\label{eq:FF0-BCL} \\
f_T (q^2) & = & 
\frac{m_B + m_\pi}{m_B \left ( 1 - \hat q_*^2 \right )} 
\sum\limits_{k = 0}^{k_{\rm max}} b^{(T)}_k (q_0^2)
\left [ z (q^2, q_0^2) \right ]^k , 
\label{eq:FFT-BCL} 
\end{eqnarray}
where $\hat q_*^2 = q^2/m_{B^*}^2$ and the function 
$z (q^2, q_0^2)$ is defined in Eq.~(\ref{eq:zqq0}).  
In this expansion the shape of the form factor is determined
by the values of~$b_k$, with truncation at $k_{\rm max} = 2$ 
or~$3$. The value of the free parameter~$q_0^2$ is proposed 
to be the so-called optimal one $q_0^2 = q_{\rm opt}^2 =  
\left ( m_B + m_\pi \right ) \left ( 
\sqrt{m_B} - \sqrt{m_\pi} \right )^2$~\cite{Bourrely:2008za},
which is obtained as the solution of the equation 
$z (0, q_0^2) = - z (m_-^2, q_0^2)$ (the latter condition 
means that the physical range $0 < q^2 \le m_-^2$ 
is projected onto a symmetric interval on the real 
axis in the complex $z$-plane). 
The prefactors $1/(1 - \hat q_*^2)$ in $f_+ (q^2)$ 
and $f_T (q^2)$ allow one to get the right asymptotic 
behavior $\sim 1/q^2$ predicted by the perturbative QCD. 
In Ref.~\cite{Becher:2005bg,Bourrely:2008za} an additional 
restriction on the series coefficients was discussed.
In particular, in the case of $f_+ (q^2)$ at $q^2 \sim m_+^2$, 
the threshold behavior of the form factor results 
in a constraint on its derivative, 
$d f_+ /dz |_{z = - 1} = 0$~\cite{Bourrely:2008za}, 
which allows one to eliminate the last term 
in the truncated expansion as follows: 
\begin{equation} 
b^{(+)}_{k_{\rm max}} (q_0^2) = 
- \frac{(-1)^{k_{\rm max}}}{k_{\rm max}}
\sum_{k = 0}^{k_{\rm max} - 1}
(-1)^k \, k \, b^{(+)}_k (q_0^2) . 
\label{eq:b-coeff-relation}  
\end{equation}
In the case of $f_0 (q^2)$ the threshold behavior 
is different and a similar relation is not applied. 
A detailed analysis of the additional constraints 
based on the threshold behavior of the tensor 
$f_T (q^2)$ form factor in the $B \to \pi$ transition 
has not yet been performed. This behavior, however, 
is not expected to be very different from the one 
found for the vector $f_+ (q^2)$ form factor. So, 
one may as well put the condition on the derivative 
$d f_T /dz |_{z = - 1} = 0$ in this case, which allows 
to eliminate the last term in the truncated expansion 
for $f_T (q^2)$. This was used in the analysis applied 
for fitting the tensor $B \to K$ transition form factor 
by the HPQCD Collaboration~\cite{Bouchard:2013eph}.


\section{\label{sec:FF-fits}
Extraction of the $f_+ (q^2)$ Form-Factor Shape
}

\subsection{\label{ssec:BF(B0-pi-ell-nu)}
The $B^0 \to \pi^- \ell^+ \nu_\ell$ Branching Fraction 
}

\begin{figure}[tb] \center
\includegraphics[width = 0.4\textwidth]{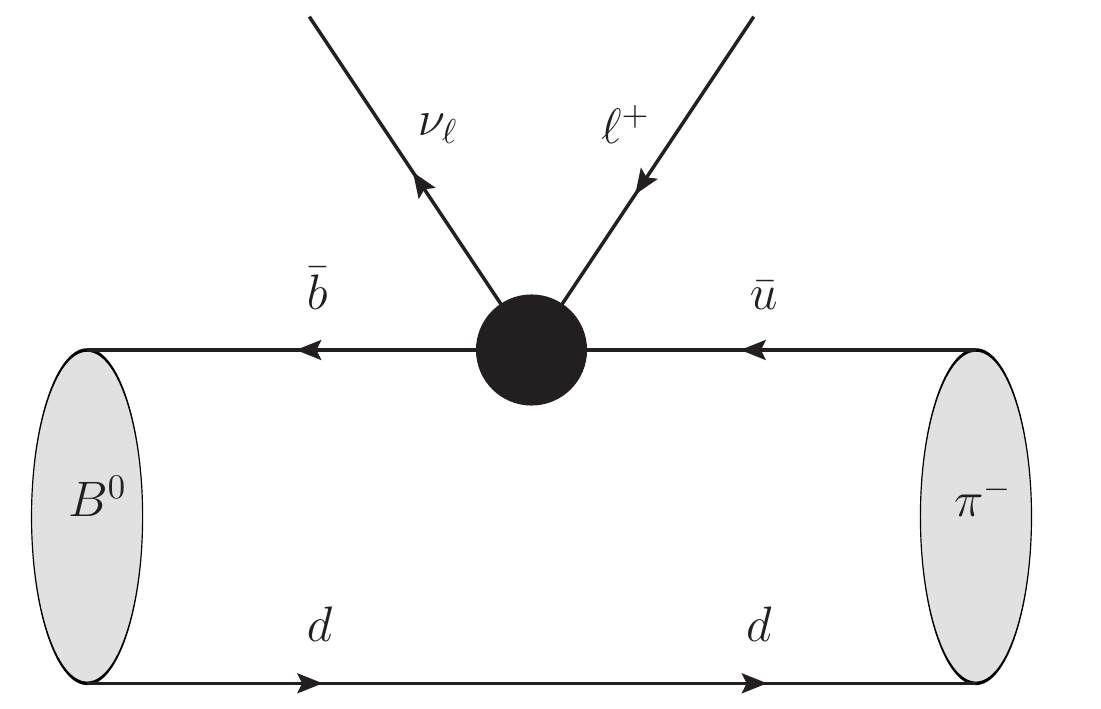} 
\caption{\label{fig:FD1}
Feynman diagram for the $B^0 \to \pi^- \ell^+ \nu_\ell$ decay.
}
\end{figure}

The charged-current Lagrangian inducing 
the $b \to u$ transition in the~SM is:
\begin{equation}
{\cal L}_W (x) = - \frac{g}{2 \sqrt 2} \, V_{ub} \left [ 
\bar u (x) \gamma_\mu \left ( 1 - \gamma_5 \right ) b (x) 
\right ] W^\mu (x)
+ \mathrm{h.\ c.} ,  
\label{eq:CKM-Lagrangian}
\end{equation}
where $g$ is the $SU(2)_L$ coupling, 
$V_{ub}$ is the element of the CKM matrix, 
$u (x)$ and $b (x)$ are the $u$- and $b$-quark fields, 
and $W (x)$ is the $W$-boson field. Feynman diagram 
for the $B^0 \to \pi^- \ell^+ \nu_\ell$ decay 
is shown in Fig.~\ref{fig:FD1} and the one 
for the $B^+ \to \pi^0 \ell^+ \nu_\ell$ decay 
differs by the exchange of the spectator-quark 
flavor ($d \to u$) only.  
The $B \to \pi$ transition matrix element  
entering the $B$-meson decay $B \to \pi \ell \nu_\ell$
can be parametrized in terms of two form factors~$f_+ (q^2)$ 
and~$f_0 (q^2)$ as follows~\cite{Neubert:1993mb,Burdman:1993es}:
%
\begin{eqnarray} 
\langle \pi (p_\pi) | \bar u \gamma^\mu b | B (p_B) \rangle 
& = & 
f_+ (q^2) \left [ 
p_B^\mu + p_\pi^\mu - \frac{m_B^2 - m_\pi^2}{q^2} \, q^\mu 
\right ] 
\nonumber \\ 
& + & 
f_0 (q^2) \, \frac{m_B^2 - m_\pi^2}{q^2} \, q^\mu .
\label{eq:Bpimatrixel}
\end{eqnarray} 
%
Here,~$p_B$ ($m_B$) and~$p_\pi$ ($m_\pi$) are the 
four-momenta (masses) of the $B$- and $\pi$-mesons, 
respectively. In the isospin-symmetry limit,  
the form factors in the charged-current 
matrix element~(\ref{eq:Bpimatrixel}) are exactly 
the same as the ones in Eq.~(\ref{eq:ME-vector}) 
in the FCNC process $B \to \pi \, \ell^+ \ell^-$. 

Measurements of the $B^0 \to \pi^- \ell^+ \nu_\ell$ 
and $B^+ \to \pi^0 \ell^+ \nu_\ell$ decays, 
where $\ell = e, \, \mu$, allow to extract 
both the CKM matrix element~$|V_{ub}|$ and 
the shape of the $f_+ (q^2)$ form factor. 
The differential branching fractions of the above 
processes can be written in the form~\cite{Beringer:1900zz}: 
%
\begin{equation}
\frac{d\Gamma (B \to \pi \ell^+ \nu_\ell)}{d q^2} = 
C_P \, \frac{G_F^2 |V_{ub}|^2}{192 \pi^3 m_B^3} \, 
\lambda^{3/2} (q^2) 
f_+^2 (q^2) , 
\label{eq:DRBpi}
\end{equation}
where $G_F$ is the Fermi constant, 
$C_P$ is the isospin factor with  
$C_P = 1$ for the $\pi^+$-meson and 
$C_P = 1/2$ for the $\pi^0$-meson,   
$\lambda (q^2)$ is the standard three-body 
kinematic factor~(\ref{eq:lambda}),  
$q^\mu = p_\ell^\mu + p_\nu^\mu$ is the total four-momentum 
transfer, bounded by $m_\ell^2 \le q^2 \le (m_B - m_\pi)^2$, 
and~$p_\ell^\mu$ and~$p_\nu^\mu$ 
are the four-momenta of the charged 
lepton and the neutrino, respectively. 
In general, the $B \to \pi$ transition matrix 
element~(\ref{eq:Bpimatrixel}) depends on two form factors. 
In practice, however, only $f_+ (q^2)$ is measurable in the 
$B \to \pi \ell \nu_\ell$ decays with $\ell=e, \mu$, 
since the contribution of the scalar form factor $f_0 (q^2)$ 
to the decay rate is suppressed by the mass ratio of 
the charged lepton to the $B$-meson~\cite{Burdman:1993es}.  

The values of~$G_F$, $m_B$, and~$m_\pi$ are known 
with high accuracy~\cite{Beringer:1900zz}, while 
the experimentally derived value of~$|V_{ub}|$ 
depends somewhat on the extraction method and 
$B$-meson decays considered. This is discussed 
at great length in the Particle Data Group (PDG) 
reviews~\cite{Beringer:1900zz}. The value quoted 
from the analysis of the exclusive $B \to \pi \ell \nu_\ell$ 
decay is listed there as $|V_{ub}| = (3.23 \pm 0.31) \times 10^{-3}$. 
On the other hand, assuming the~SM, the CKM unitarity fits yield 
a value of $|V_{ub}|$ which is consistent with the previous value, 
but is about a factor~2 more precise~\cite{Beringer:1900zz}: 
$|V_{ub}| = (3.51^{+0.15}_{-0.14}) \times 10^{-3}$,
which we use as our default value in the numerical estimates.

The partial branching fractions for 
the $B^0 \to \pi^- \ell^+ \nu_\ell$ decays has been 
measured by the CLEO, BaBar and Belle collaborations, 
and for the $B^+ \to \pi^0 \ell^+ \nu_\ell$ decays 
by the Belle Collaboration.
Below we give the total branching fraction 
of the $B^0 \to \pi^- \ell^+ \nu_\ell$ decay 
taking into account the recent data 
from the BaBar and Belle collaborations%
~\cite{delAmoSanchez:2010zd,Lees:2012vv,Ha:2010rf,Sibidanov:2013rkk}: 
\begin{equation}
\begin{split}
& 
{\cal B} (B^0 \to \pi^- \ell^+ \nu_\ell) \\ 
&
= \left \{
\begin{array}{rl}
\! (1.42 \pm 0.05_{\rm stat} \pm 0.07_{\rm syst}) \times 10^{-4} & 
        [\mbox{BaBar, 2011}]~, \\
\! (1.45 \pm 0.04_{\rm stat} \pm 0.06_{\rm syst}) \times 10^{-4} & 
        [\mbox{BaBar, 2012}]~, \\
\! (1.49 \pm 0.04_{\rm stat} \pm 0.07_{\rm syst}) \times 10^{-4} & 
        [\mbox{Belle, 2011}]~, \\   
\! (1.49 \pm 0.09_{\rm stat} \pm 0.07_{\rm syst}) \times 10^{-4} & 
        [\mbox{Belle, 2013}]~.   
\end{array} 
\right. 
\end{split}
\label{eq:Bo-pim-el-nu-data}
\end{equation}
All these measurements are in excellent agreement with each other, 
and with the one for the $B^+ \to \pi^0 \ell^+ \nu_\ell$ decay 
reported by the Belle Collaboration~\cite{Sibidanov:2013rkk}: 
\begin{equation}
{\cal B} (B^+ \to \pi^0 \ell^+ \nu_\ell) = 
(0.80 \pm 0.08_{\rm stat} \pm 0.04_{\rm syst}) \times 10^{-4} .  
\label{eq:Bp-pi0-el-nu-data}
\end{equation}
Both the collaborations have presented differential 
distributions in~$q^2$ relevant for the extraction 
of $f_+ (q^2)$ from data~\cite{delAmoSanchez:2010zd,%
Lees:2012vv,Ha:2010rf,Sibidanov:2013rkk}.  
We show them in the next subsection, where also our 
fitting procedure is presented. 


\subsection{\label{ssec:FF-fits-method}
Fitting Procedure
}

In this subsection the extraction of the $f_+ (q^2)$  
form-factor shape from the dilepton invariant-mass 
spectra in the $B^0 \to \pi^- \ell^+ \nu_\ell$ and 
$B^+ \to \pi^0 \ell^+ \nu_\ell$ decays measured by 
the BaBar~\cite{delAmoSanchez:2010zd,Lees:2012vv} 
and Belle~\cite{Ha:2010rf,Sibidanov:2013rkk} 
collaborations is explained.  
All four $f_+ (q^2)$ form-factor parametrizations 
from Sec.~\ref{sec:FF-parametrizations} are examined 
to test their consistency with the experiment 
in terms of the best-fit values resulting from the 
$\chi^2$-distribution function~\cite{Beringer:1900zz}. 

\begin{figure*}[tb] 
\includegraphics[width = 0.45\textwidth]{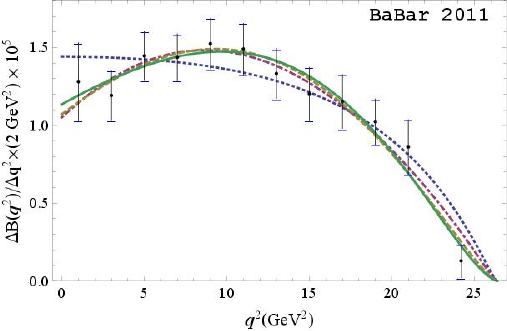} 
\hfill 
\includegraphics[width = 0.45\textwidth]{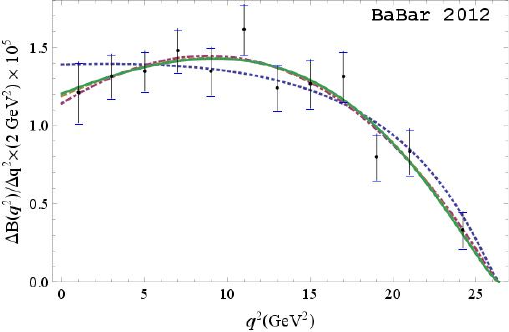} 
\\[3mm] 
\includegraphics[width = 0.45\textwidth]{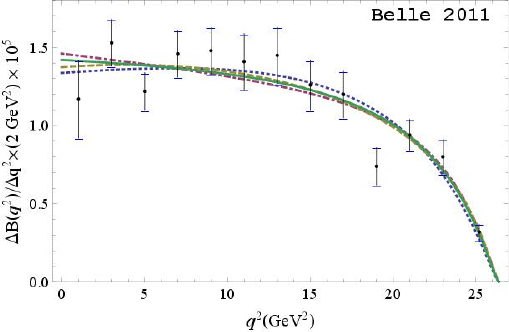} 
\hfill 
\includegraphics[width = 0.45\textwidth]{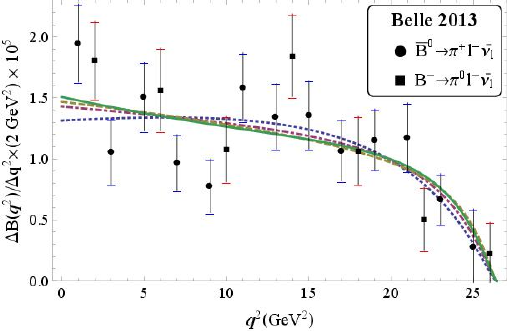} 
\caption{\label{fig:BRshape}   
(Color online.) 
Partial $\Delta {\cal B} (q^2)/\Delta q^2$ spectra 
for the $B^0 \to \pi^- \ell^+ \nu_\ell$ and 
$B^+ \to \pi^0 \ell^+ \nu_\ell$ decays, where $\ell = e,\, \mu$. 
The data points (black dots and squares) are placed 
in the middle of each bin.
The error bars (blue) include the total experimental uncertainties. 
The curves show the results of the fit to the data for  
the four form-factor parametrizations discussed in the text: 
BK~(\ref{eq:FFp-BK}) (thick dotted blue line),
BZ~(\ref{eq:FFp-BZ}) (thick dashed purple line),
BGL~(\ref{eq:FF-BGL}) with $k_{\rm max} = 2$ (thick dot-dashed yellow line), 
and BCL~(\ref{eq:FFp-BCL}) with $k_{\rm max} = 2$ (thick solid green line).   
The upper-left and upper-right plots correspond 
to the BaBar 2011~\cite{delAmoSanchez:2010zd} 
and 2012~\cite{Lees:2012vv} data sets,
while the lower-left and lower-right plots are plotted 
based on the Belle 2011~\cite{Ha:2010rf} 
and 2013~\cite{Sibidanov:2013rkk} data sets. 
} 
\end{figure*}

\begin{figure*}[tb] 
\begin{center}
\includegraphics[width = 0.45\textwidth]{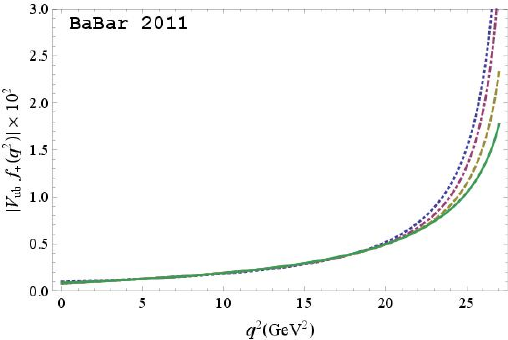} 
\hfill 
\includegraphics[width = 0.45\textwidth]{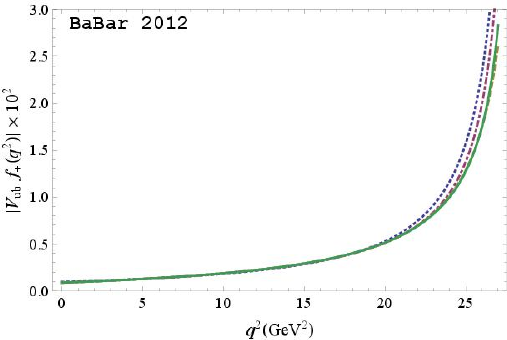} 
\\[3mm]
\includegraphics[width = 0.45\textwidth]{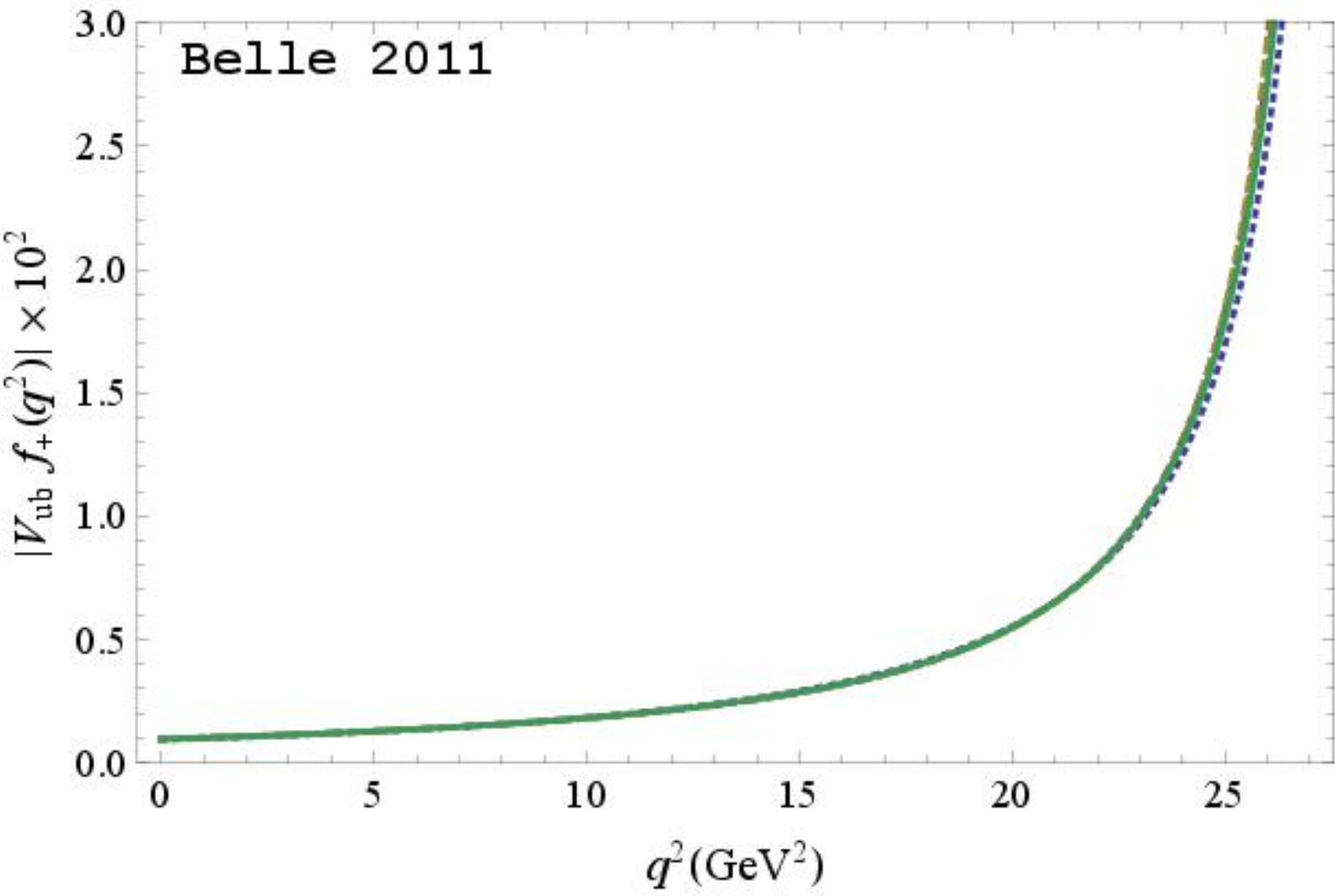} 
\hfill 
\includegraphics[width = 0.45\textwidth]{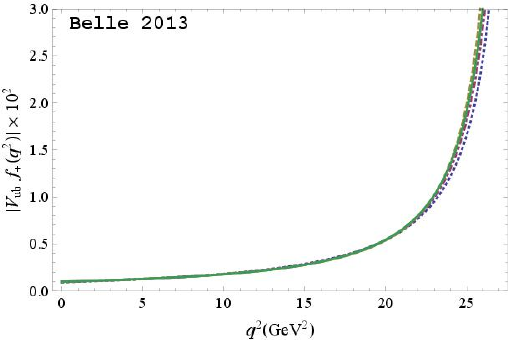} 
\end{center}
\caption{\label{fig:FFshape}  
(Color online.) 
The $f_+ (q^2)$ form-factor shapes
in the decay $B \to \pi \ell \nu_\ell$ multiplied 
by the CKM matrix element~$|V_{ub}|$ following 
from the BaBar~\cite{delAmoSanchez:2010zd,Lees:2012vv} 
and Belle~\cite{Ha:2010rf,Sibidanov:2013rkk} data. 
The curves show the results of the fit to these data:  
BK~(\ref{eq:FFp-BK}) (thick dotted blue line),
BZ~(\ref{eq:FFp-BZ}) (thick dashed purple line),
BGL~(\ref{eq:FF-BGL}) with $k_{\rm max} = 2$ 
(thick dot-dashed yellow line), 
and BCL~(\ref{eq:FFp-BCL}) with $k_{\rm max} = 2$ 
(thick solid green line) parametrizations.
}
\end{figure*}

The fitted form factor is presented as a function 
of~$q^2$ which contains a set of~$k$ unknown 
parameters $\alpha_1, \ldots, \alpha_k$:
\begin{equation}
f_+ (q^2) = f (q^2; \alpha_1, \ldots, \alpha_k) .
\label{eq:FFsetpar}
\end{equation}
Given the experimental values~$y_i$ of the partial 
branching fractions $\Delta {\cal B} (q^2)/\Delta q^2$ 
in bins of~$q^2$, with their uncertainties~$\sigma_i$, 
the $\chi^2$-distribution function is defined as 
follows~\cite{Beringer:1900zz}:  
\begin{equation}
\chi^2 = \sum\limits_{i=1}^N \frac{
\left [ y_i - F (x_i; \alpha_1, \ldots, \alpha_k) \right ]^2
}{\sigma_i^2} ,
\label{eq:chi-sq}
\end{equation}
where~$N$ is the number of experimental points 
and $F (x_i; \alpha_1, \ldots, \alpha_k)$ denotes 
the theoretical estimates of the partial branching 
fractions $\Delta {\cal B} (q^2)/\Delta q^2$
for the given parametrization:
\begin{equation}
F (x_i; \alpha_1, \ldots, \alpha_k) = 
\int\limits_{x_i - a_i/2}^{x_i + a_i/2} \!
\frac{d {\cal B} (q^2)}{dq^2} \, dq^2 ,
\label{eq:Ftheor}
\end{equation}
%
with~$x_i$ and~$a_i$ being the center and the width of the $i$th bin.  
The standard minimization procedure of the $\chi^2$-function 
(minimum of this function is denoted as $\chi^2_{\rm min}$)
allows us to extract the values of fitted parameters 
$\alpha_{1, {\rm min}}, \ldots, \alpha_{k, {\rm min}}$, 
which are considered to be their best-fit values.
The results obtained by using the four form-factor 
parametrizations for different sets of experimental 
data obtained by the BaBar~\cite{delAmoSanchez:2010zd,Lees:2012vv} 
and Belle~\cite{Ha:2010rf,Sibidanov:2013rkk} collaborations 
are presented in Figs.~\ref{fig:BRshape} and~\ref{fig:FFshape}, 
respectively, and the numerical values for $\chi_{\rm min}^2$/ndf, 
where ndf is the number of degrees of freedom, 
and the corresponding $p$-values are presented 
in Table~\ref{tab:chisqvalues}. 
In this analysis we have assumed that the experimental 
points are all uncorrelated.



\begin{table*}[tb]
\caption{\label{tab:chisqvalues}
Summary of the $\chi_{\rm min}^2$/ndf values, 
where ndf is the number of degrees of freedom,
(corresponding $p$-values)  
for different sets of experimental data (rows) 
and four form-factor parametrizations discussed in the text (columns).
} 
%
%
\begin{center}
\begin{tabular}{|l|c|c|c|c|}
     \hline 
& BK \cite{Becirevic:1999kt} 
& BZ \cite{Ball:2004ye}  
& BGL \cite{Boyd:1995sq} 
& BCL \cite{Bourrely:2008za} 
      \\ \hline
BaBar 2011 \cite{delAmoSanchez:2010zd} &  
9.93/10 (45\%)   &  4.80/9 (85\%)  &  4.12/9 (90\%)   &   3.75/9 (93\%) \\
      \hline
BaBar 2012 \cite{Lees:2012vv} &  
8.68/10 (56\%)   &  5.50/9 (79\%)  &  5.65/9 (77\%)   &   5.73/9 (77\%) \\
      \hline
Belle 2011 \cite{Ha:2010rf} &  
15.86/11 (15\%)  & 14.55/10 (15\%) &  12.97/10 (23\%) & 14.44/10 (15\%) \\
      \hline
Belle 2013 \cite{Sibidanov:2013rkk} &  
24.41/18 (14\%)  & 23.55/17 (13\%) &  24.16/17 (12\%) & 23.26/17 (14\%) \\
      \hline
BaBar \& Belle  &  
44.99/43 (39\%)  & 44.91/42 (35\%) &  44.56/42 (36\%) & 44.77/42 (36\%) \\
      \hline 
\end{tabular}
\end{center}
\end{table*}

From Table~\ref{tab:chisqvalues} it follows that the 
smallest value for $\chi_{\rm min}^2$/ndf corresponds 
to the simplest Becirevic-Kaidalov parametrization. 
From the rest of the specified parametrizations, 
the Boyd-Grinstein-Lebed one has the smallest 
$\chi_{\rm min}^2$/ndf value and we will use it for all 
the form factors entering the $B \to \pi \ell^+ \ell^-$ decay. 

\begin{figure*}[tb] 
\begin{center}
\includegraphics[width = 0.40\textwidth]{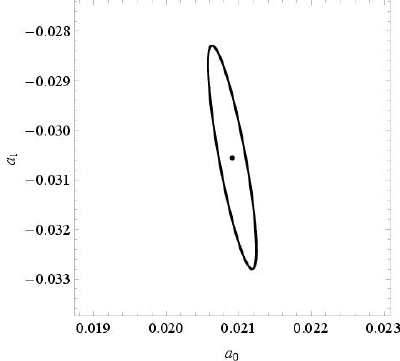} 
\hfill 
\includegraphics[width = 0.40\textwidth]{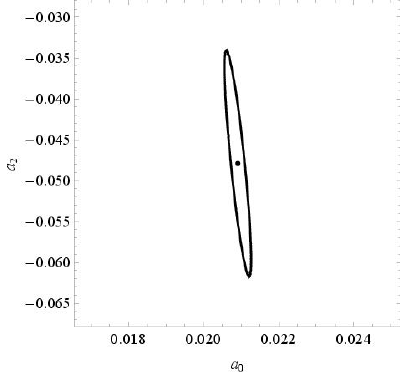}  
\\[3mm]
\includegraphics[width = 0.40\textwidth]{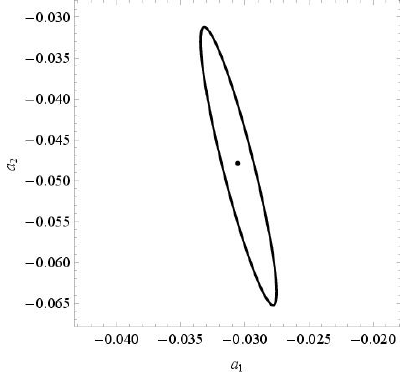} 
\hfill 
\includegraphics[width = 0.40\textwidth]{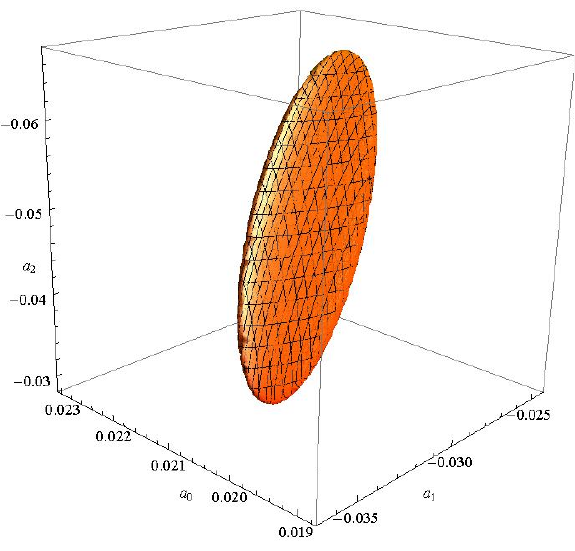} 
\end{center}
\caption{\label{fig:FFp-correlations}  
(Color online.) 
The two-dimensional correlations among 
the fitted parameters~$a_0$, $a_1$ and~$a_2$  
entering the BGL-parametrization 
of the form factor $f_+ (q^2)$: 
$a_0 - a_1$ (upper-left plot), 
$a_0 - a_2$ (upper-right plot) and 
$a_1 - a_2$ (lower-left plot).  
The three-dimensional correlation among all three 
fitted parameters is shown in the lower-right plot.  
}
\end{figure*}

The combined analysis of the BaBar and Belle data 
yields the following set of fitted parameters 
entering the $f_+ (q^2)$ form factor expansion in 
the BGL parametrization, truncated at $k_{\rm max} = 2$: 
\begin{eqnarray}
a_0 & = & \hphantom{-}0.0209 \pm 0.0004 , 
\nonumber \\ 
a_1 & = &           - 0.0306 \pm 0.0031 , 
\label{eq:a-fitted} \\ 
a_2 & = &           - 0.0473 \pm 0.0189 .  
\nonumber 
\end{eqnarray}
The extracted numerical values depend 
on the CKM matrix element~$|V_{ub}|$ and 
correspond to the PDG value~\cite{Beringer:1900zz}:   
$|V_{ub}| = (3.51^{+0.15}_{-0.14}) \times 10^{-3}$. 
The errors specified in the coefficients~(\ref{eq:a-fitted}) 
are the square roots of the covariance matrix~$U_{ij}$ 
for the BGL form-factor coefficients which can be derived 
from the $\chi^2$-function~(\ref{eq:chi-sq}) 
as follows~\cite{Beringer:1900zz}: 
\begin{equation}  
\left ( U^{-1} \right )_{ij} = \frac{1}{2} \left. 
\frac{\partial^2 \chi^2}{\partial\alpha_i \partial\alpha_j} 
\right |_{\alpha_k = \hat\alpha_k} ,
\label{eq:cov-matrix}
\end{equation}
where $\hat\alpha_k$ are the best-fit values 
of the fitting parameters. The  function 
$F (x_i; \alpha_1, \ldots, \alpha_k)$
in the BGL form factor depends 
linearly on the unknown parameters, which simplifies 
the analysis. The corresponding correlation 
matrix~$r_{ij}$ is connected with 
the covariance matrix by the relation 
$r_{ij} = U_{ij}/(\sigma_i \, \sigma_j)$,  
where $\sigma_i^2$ is the variance of~$\alpha_i$. 
For the BGL form factor with the truncation 
at $k_{\rm max} = 2$, the following 
$(3 \times 3)$ correlation matrix was obtained: 
\begin{equation} 
r_{ij} = \left ( 
\begin{array}{ccc} 
    1 & -0.26 & -0.43 \\ 
-0.26 &     1 & -0.68 \\ 
-0.43 & -0.68 &     1
\end{array}
\right ) .
\label{eq:a-correlation}
\end{equation}
One can see the sizable 
correlation of the third coefficient~$a_2$ 
in the $z$-expansion with the other two~$a_0$ 
and~$a_1$. This is shown in 
Fig.~\ref{fig:FFp-correlations}. The relative 
error on the coefficient~$a_2$ is approximately~40\%
as can also be seen in Eq.~(\ref{eq:a-fitted}). 

\begin{figure}[tb]  
\begin{center}
\includegraphics[width = 0.40\textwidth]{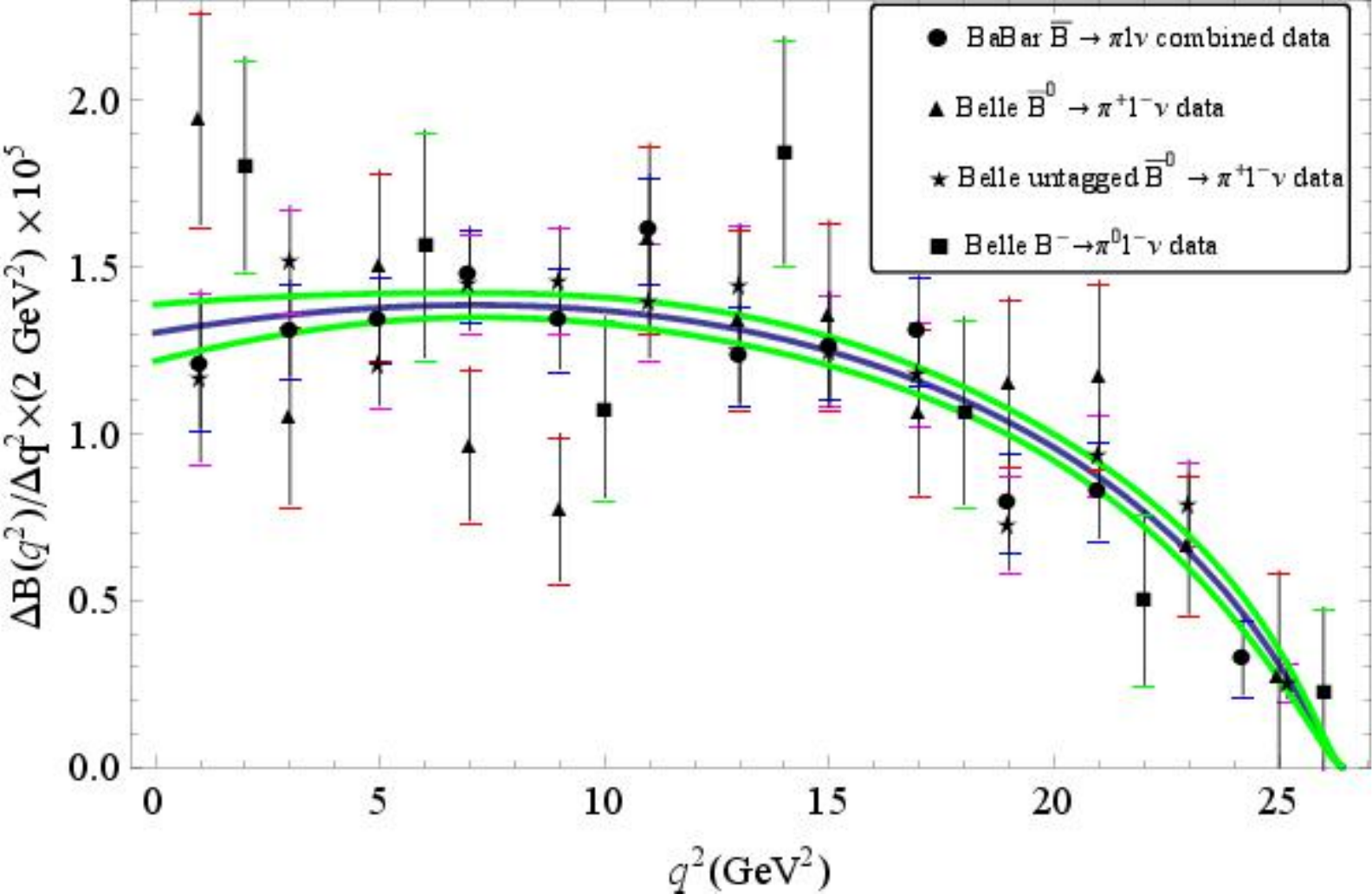} \\
\includegraphics[width = 0.40\textwidth]{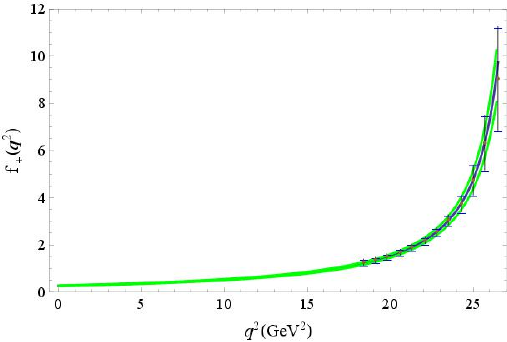} 
\end{center}
\caption{\label{fig:BR-FF-shapeComb}   
(Color online.) 
Partial $\Delta {\cal B} (q^2)/\Delta q^2$ spectra  
for the decays $B^0 \to \pi^- \ell^+ \nu_\ell$ and 
$B^+ \to \pi^0 \ell^+ \nu_\ell$ are 
presented on the upper plot. 
The $f_+ (q^2)$ form factor is shown on the lower plot.  
The BGL parametrization is adopted as the preferred choice. 
Results are obtained by combining the experimental 
data by the BaBar~\cite{Lees:2012vv} and 
Belle~\cite{Ha:2010rf,Sibidanov:2013rkk} 
collaborations and, in addition, the value 
$|V_{ub}| = (3.51^{+0.15}_{-0.14}) \times 10^{-3}$%
~\cite{Beringer:1900zz} is used to extract 
explicitly the form-factor shape. The existing 
Lattice-QCD data~\cite{Dalgic:2006dt} on the 
form factor are presented as the vertical bars 
on the lower plot.    
} 
\end{figure}

The results from the combined analysis 
of the BaBar~\cite{Lees:2012vv} and 
Belle~\cite{Ha:2010rf,Sibidanov:2013rkk} 
data sets are shown in Fig.~\ref{fig:BR-FF-shapeComb} (upper plot). 
Following the numerical analysis presented above,  
the resulting shape of the $f_+ (q^2)$ form 
factor is presented on the lower plot in 
Fig.~\ref{fig:BR-FF-shapeComb}, using the  
BGL parametrization  and the PDG value 
$|V_{ub}| = (3.51^{+0.15}_{-0.14}) \times 10^{-3}$%
~\cite{Beringer:1900zz}. 
The existing Lattice-QCD results~\cite{Dalgic:2006dt} 
on the $f_+ (q^2)$ form factor are presented 
as vertical bars on the lower plot in Fig.~\ref{fig:BR-FF-shapeComb}, which are
in good agreement with our estimate of the same in the overlapping $q^2$-region
 (within the uncertainties of the lattice data, as indicated).


\section{\label{sec:FF0-T-fits} 
Determination of $f_0 (q^2)$ and $f_T (q^2)$ Shapes
}

As pointed out earlier, the form factor~$f_0 (q^2)$ 
is not required for either the charged-current decay 
$B \to \pi \ell \nu_\ell$ or the FCNC semileptonic 
$B \to \pi \ell^+ \ell^-$ decay with $\ell = e, \mu$, 
as its contribution to the branching fraction is 
suppressed by the smallness of the lepton mass squared. 
However, for the sake of completeness involving the 
semileptonic processes with $\ell^\pm = \tau^\pm$, 
we also work out the $f_0 (q^2)$ form factor. 
The information on the form factors $f_+ (q^2)$ and 
$f_0 (q^2)$ for the $B \to \pi$ and $B \to K$ transitions 
is available, though the lattice results on the $B \to \pi$ 
form factor $f_T (q^2)$ are still scant. For our analysis,  
we use an Ansatz for the $SU(3)_F$-symmetry breaking 
to obtain the shape of~$f^{B\pi}_T (q^2)$ from the 
corresponding $B \to K$ form factor $f^{BK}_T (q^2)$. 
We show subsequently that our Ansatz, which assumes 
that the $SU(3)_F$-symmetry breaking in $f_T (q^2)$ 
is an average of the corresponding symmetry-breaking 
effects in the form factors $f_+ (q^2)$ and $f_0 (q^2)$,
yields an $f^{B\pi}_T (q^2)$, which is in good agreement 
with the preliminary results on this form factor, 
obtained from lattice in the low-recoil region.

\subsection{\label{ssec:f0-FF}  
The $f_0 (q^2)$ Form Factor
}

\begin{figure}[tb] 
\begin{center} 
\includegraphics[width = 0.45\textwidth]{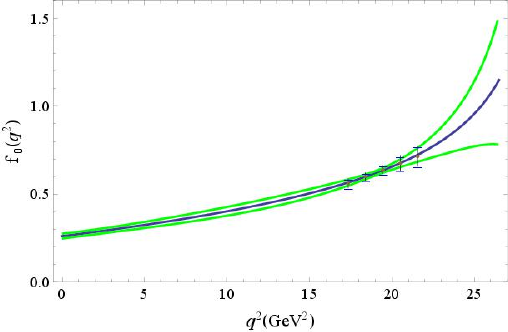}  
\end{center} 
\caption{\label{fig:scalar-FF} 
(Color online.)
The scalar $B \to \pi$ transition form factor 
$f_0 (q^2)$ in the entire kinematic region 
using the BGL parametrization. The solid 
green lines show the uncertainty in the form 
factor. The vertical bars are the Lattice-QCD 
data~\cite{Dalgic:2006dt} used for fixing the 
form-factor shape.   
}
\end{figure}

The parameters in $f_0 (q^2)$ can be obtained 
from the existing results of the $B \to \pi$  
transition form factor calculated by the HPQCD 
Collaboration~\cite{Dalgic:2006dt}. In addition 
we use the exact relation between $f_+ (q^2)$ 
and $f_0 (q^2)$ at $q^2 = 0$: 
\begin{equation} 
f_+ (0) = f_0 (0) ,  
\label{eq:FFp-FF0-zero}  
\end{equation} 
which follows from the requirement of the 
finiteness of the $B \to \pi$ transition 
matrix element~(\ref{eq:ME-vector}) at this 
point. To fix~$f_0 (0)$, we use the reference
point $f_+ (0) = 0.261 \pm 0.014$, extracted 
by us from the experimental data. The form-factor 
parametrization we use for $f_0 (q^2)$ 
follows our default choice from the analysis 
of $f_+ (q^2)$~--- the BGL expansion 
in~$z (q^2, q_0^2)$ truncated at $k_{\rm max} = 2$. 
The set of the fitted parameters entering 
$f_0 (q^2)$ is as follows:  
\begin{eqnarray}
a_0 & = & \hphantom{-}0.0201 \pm 0.0007 , 
\nonumber \\ 
a_1 & = &           - 0.0394 \pm 0.0096 , 
\label{eq:a-f0-fitted} \\ 
a_2 & = &           - 0.0355 \pm 0.0556 , 
\nonumber 
\end{eqnarray}
and the  
correlation matrix ($i, j = 1, 2, 3$) is:  
\begin{equation} 
r_{ij} = \left ( 
\begin{array}{ccc} 
    1 &  0.72 & -0.82 \\ 
 0.72 &     1 & -0.96 \\ 
-0.82 & -0.96 &     1
\end{array}
\right ) .
\label{eq:a-f0-correlation}
\end{equation}
One sees strong correlations among 
all the fitted parameters, which can be 
well approximated as linear.  
 The resulting shape 
is shown in Fig.~\ref{fig:scalar-FF}. The solid (green) 
lines specify the from-factor uncertainty which grows 
with increasing~$q^2$. This trend is reflected also in 
the lattice data~\cite{Dalgic:2006dt} 
(shown by the vertical bars in Fig.~\ref{fig:scalar-FF}).

\subsection{\label{ssec:fT-FF}  
The $f_T (q^2)$ Form Factor
} 

\begin{figure}[tb] 
\begin{center} 
\includegraphics[width = 0.45\textwidth]{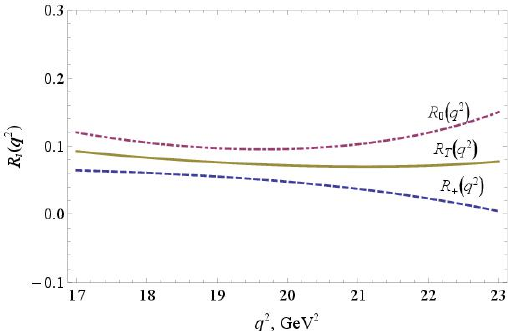} 
\hfill 
\includegraphics[width = 0.45\textwidth]{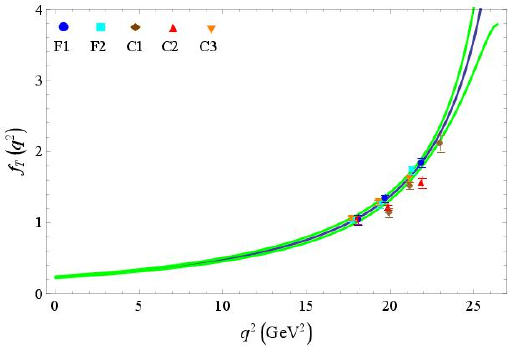}  
\end{center} 
\caption{\label{fig:tensor-FF}
(Color online.)  
The $SU(3)_F$-symmetry breaking functions $R_+ (q^2)$, 
$R_0 (q^2)$ and $R_T (q^2)$ (the upper plot)  
in the $q^2$-range accessible by the Lattice-QCD 
simulations and the tensor $B \to \pi$ transition 
form factor $f_T (q^2)$ (the lower plot)  
in the entire kinematic region. The sets of vertical bars 
in the large-$q^2$ region are the preliminary results 
from the HPQCD Collaboration~\cite{Bouchard:2013zda} 
presented at the Lattice-2013 Conference.  
The legend on the lower plot specifies the 
lattice ensembles as used in the $B \to K$ 
transitions~\cite{Bouchard:2013eph}  
by the HPQCD Collaboration.  
}
\end{figure}

As already mentioned, there is at present only scant 
information from the lattice on  the $B \to \pi$ tensor form factor 
$f_T^{B\pi} (q^2)$. 
So, one needs to find a reliable method to extract 
it from the existing model-independent data. We use 
an $SU(3)_F$-symmetry breaking Ansatz involving both  
the $B \to K$ and $B \to \pi$ transition form factors. 
We recall that all three $B \to K$ transition form 
factors $f_+^{BK} (q^2)$, $f_0^{BK} (q^2)$ and 
$f_T^{BK} (q^2)$ have been calculated recently by the 
HPQCD Collaboration~\cite{Bouchard:2013eph,Bouchard:2013mia}  
and the two $B \to \pi$ transition form factors 
$f_+^{B\pi} (q^2)$ and $f_0^{B\pi} (q^2)$ are 
also known~\cite{Dalgic:2006dt}. Of course, lattice 
results are available only in the small-recoil limit. 
With this at hand, we first estimate the 
$SU(3)_F$-symmetry-breaking corrections in the already 
known vector and scalar form factors and use these  
corrections to estimate the $B \to \pi$ tensor 
form factor $f_T^{B\pi} (q^2)$ from the corresponding 
$B \to K$ transition form factor $f_T^{BK} (q^2)$. 
We introduce the following measures 
of the $SU(3)_F$-symmetry breaking corrections 
in the transition form factors:  
\begin{equation}
R_i (q^2) = \frac{f_i^{BK} (q^2)}{f_i^{B\pi} (q^2)} - 1 , 
\label{eq:FF-ratio}
\end{equation}
where $i = +,\, 0,\, T$. 
The curves for the $SU(3)_F$-symmetry 
breaking functions $R_+ (q^2)$ and $R_0 (q^2)$, 
calculated for the central values of the form 
factors from the lattice for the small-recoil 
region, are presented in the upper plot 
in Fig.~\ref{fig:tensor-FF}.  
As expected, breaking effects of order~10\% 
are seen in both the ratios. We expect 
that the $SU(3)_F$-symmetry breaking effect 
in the third ratio, $R_T (q^2)$, is of the 
same order. For the sake of definiteness, 
we assume that the ratio $R_T (q^2)$ of the 
tensor form factors is the average of the 
other two: $R_+ (q^2)$ and $R_0 (q^2)$, 
\begin{equation}
R_T (q^2) = \frac{1}{2} 
\left [ R_+ (q^2) + R_0 (q^2) \right ] .  
\label{eq:FFT-ratio-def}
\end{equation}
We estimate the accuracy of this relation 
in the low-$q^2$ region, where the methods 
based on HQS (and its leading-order breakings)
can be gainfully used to quantify it 
(see Sec.~\ref{ssec:ansatz-accuracy} for details).
We expect that this relation holds to a good extent 
in the remaining large-$q^2$ region, and estimate 
the associated uncertainty to be about~5\%.  
The corresponding function $R_T (q^2)$  
is presented in the upper plot in Fig.~\ref{fig:tensor-FF} 
as the central curve. Explicit values of 
this function in the small-recoil region 
are presented in Table~\ref{tab:FFT-values}. 
The errors reflect the uncertainties of 
the lattice calculations and we assume that 
the errors in the $B \to \pi$ and $B \to K$ 
transition form factors are uncorrelated. 

The values of the $f_T^{B\pi} (q^2)$ form 
factor are then obtained by rescaling them 
from the known values of the $f_T^{BK} (q^2)$ 
form factor~\cite{Bouchard:2013eph} 
by utilizing the relation:
\begin{equation}
f_T^{B\pi} (q^2) = \frac{f_T^{BK} (q^2)}{1 + R_T (q^2)}~.  
\label{eq:FFT-values}  
\end{equation}
They are presented in Table~\ref{tab:FFT-values}.  
The variance of $f_T^{B \pi} (q^2)$ is calculated 
by adding the errors of $f_T^{BK} (q^2)$ and $R_T (q^2)$ 
in quadrature. The normalization at $q^2 = 0$: 
$f_T^{B\pi} (0) = 0.231 \pm 0.013$, which results 
from the value $f_+^{B\pi} (0) = 0.261 \pm 0.014$, 
extracted by us from the experimental data 
on the $B \to \pi \ell \nu_\ell$ decays, 
and the heavy-quark symmetry relation between 
the form factors in the large-recoil limit of the 
$\pi$-meson~\cite{Charles:1998dr,Beneke:2000wa}: 
$f^{B\pi}_T (0) = \left ( 1 + m_\pi/m_B \right ) f^{B\pi}_+ (0)$. 
With all this at hand, we have a fairly constrained 
model for the $f_T^{B\pi} (q^2)$ form factor.

\begingroup 
\squeezetable 
\begin{table}[tb]
\caption{\label{tab:FFT-values}
Values of the tensor form factor $f_T^{B\pi} (q^2)$ 
at the indicated values of~$q^2$
obtained from the existing Lattice-QCD  
data on the $f_T^{BK} (q^2)$ transition 
form factor~\cite{Bouchard:2013eph} and 
the $SU(3)_F$-symmetry breaking function 
$R_T (q^2)$ defined in  
Eqs.~(\ref{eq:FF-ratio}) and~(\ref{eq:FFT-ratio-def}). 
The variance of $f_T^{B \pi} (q^2)$ 
is calculated by adding the errors of  
$f_T^{BK} (q^2)$ and $R_T (q^2)$ in quadrature.   
}
%
%
\begin{center}
\begin{tabular}{|l|c|c|c|c|}
\hline 
$q^2$, GeV$^2$  &   
18.4 &  19.1 &  19.8 &  20.6 \\ \hline
$f_T^{B K} (q^2)$ & 
$1.197 \pm 0.047$ & $1.307 \pm 0.051$ & $1.434 \pm 0.057$ & $1.608 \pm 0.069$ \\
$R_T (q^2)$  & 
$0.080 \pm 0.021$ & $0.076 \pm 0.021$ & $0.073 \pm 0.023$ & $0.071 \pm 0.023$ \\
$f_T^{B \pi} (q^2)$  & 
$1.108 \pm 0.126$ & $1.215 \pm 0.115$ & $1.337 \pm 0.117$ & $1.503 \pm 0.123$ \\ 
\hline
\hline 
$q^2$, GeV$^2$  &  
21.3 &  22.1 &  22.8 &  23.5 \\ \hline
$f_T^{B K} (q^2)$ & 
$1.793 \pm 0.082$ & $2.054 \pm 0.106$ & $2.342 \pm 0.135$ & $2.713 \pm 0.176$ \\
$R_T (q^2)$  & 
$0.070 \pm 0.037$ & $0.072 \pm 0.050$ & $0.076 \pm 0.067$ & $0.083 \pm 0.090$ \\
$f_T^{B \pi} (q^2)$  & 
$1.675 \pm 0.144$ & $1.916 \pm 0.169$ & $2.178 \pm 0.211$ & $2.506 \pm 0.302$ \\ 
\hline
\end{tabular}
\end{center}
\end{table}
\endgroup 

For the BGL parametrization of the $f_T^{B\pi} (q^2)$ 
form factor, the fitted parameters entering 
the expansion in $z (q^2, q_0^2)$ and truncated 
at $k_{\rm max} = 2$ are as follows: 
\begin{eqnarray}
a_0 & = & \hphantom{-}0.0458 \pm 0.0027 , 
\nonumber \\ 
a_1 & = &           - 0.0234 \pm 0.0124 , 
\label{eq:a-fT-fitted} \\ 
a_2 & = &           - 0.2103 \pm 0.1052 , 
\nonumber 
\end{eqnarray}
with the corresponding correlation matrix ($i, j = 1, 2, 3$): 
\begin{equation} 
r_{ij} = \left ( 
\begin{array}{ccc} 
    1 &  0.68 & -0.90 \\ 
 0.68 &     1 & -0.83 \\ 
-0.90 & -0.83 &     1
\end{array}
\right ) .
\label{eq:a-fT-correlation}
\end{equation}
Strong correlations among the fitted parameters 
are observed similar to the case of $f_0^{B\pi} (q^2)$. 

The resulting $f_T^{B \pi} (q^2)$ form factor 
is shown in the lower plot in Fig.~\ref{fig:tensor-FF}. 
Recent preliminary results~\footnote{
They were presented by C.~Bouchard {\it et al.} at the 
Lattice-2013 Conference, held recently in Mainz (Germany).} 
for this form factor at large~$q^2$ from 
the HPQCD Collaboration~\cite{Bouchard:2013zda} 
are also presented in this figure. 
The symbols (F1, F2, C1, C2, C3) and the corresponding
lattice-data points denote the various lattice ensembles 
used by this collaboration for performing the numerical 
simulations, which are the same as the ones used 
in the calculation of the $B \to K$ transition form 
factors~\cite{Bouchard:2013mia,Bouchard:2013eph}, 
namely the MILC $N_f = 2 + 1$ asqtad gauge configurations. 
Good agreement of the lattice data on $f_T^{B\pi} (q^2)$ 
in the large-$q^2$ region with our results based  
on using the $SU(3)_F$-symmetry breaking Ansatz  
is evident in this figure. 
  
As all the form factors in the $B \to \pi$ transition 
are now determined, using data and the Lattice QCD, 
we can now make model-independent predictions for 
the short-distance part of the dilepton invariant-mass 
spectrum and the decay width in the semileptonic 
$B \to \pi \, \ell^+ \ell^-$ decays. As the long-distance 
effects dominate in the resonant regions 
(such as of the $J/\psi$- and $\psi (2S)$-mesons), 
which at present are not precisely calculable, 
a sharper contrast of the SM predictions and data 
is obtained in limited regions of~$q^2$, 
which we present in subsequent sections.

\section{\label{sec:B-pi-mu-mu-HQS-limit} 
$B^+ \to \pi^+ \ell^+ \ell^-$ Decay in Low-$q^2$ Region 
}

\subsection{\label{ssec:HQS-limit}
HQS Limit
} 

As discussed in the Introduction, one can apply 
the heavy-quark symmetry techniques 
to relate the form factor $f_T (q^2)$ 
in $B^\pm \to \pi^\pm \ell^+ \ell^-$ to 
the measured form factor $f_+ (q^2)$ 
in the charged-current decay $B \to \pi \ell \nu_\ell$, 
in the large-recoil (or low-$q^2$) region. 
As shown in Ref.~\cite{Beneke:2000wa}, 
in the HQS limit (i.\,e., without taking 
into account symmetry-breaking corrections), 
$f_0 (q^2)$ and~$f_T (q^2)$ are proportional 
to $f_+ (q^2)$:
\begin{eqnarray}
& & f_0 (q^2) = \frac{m_B^2 - q^2}{m_B^2} \, f_+ (q^2) , 
\label{eq:fp-f0-relation} \\
& & f_T (q^2) = \frac{m_B + m_\pi}{m_B} \, f_+ (q^2) .
\label{eq:fp-fT-relation}
\end{eqnarray}
In the HQS limit, there is only one independent form 
factor $f_+ (q^2)$, the shape of which can be extracted 
from the analysis of the $B^0 \to \pi^- \ell^+ \nu_\ell$ 
and $B^+ \to \pi^0 \ell^+ \nu_\ell$, which we presented 
in Sec.~\ref{sec:FF-fits}.  
The decay rate of $B^+ \to \pi^+ \ell^+ \ell^-$ 
in the HQS limit is greatly simplified and takes the form:
%
\begin{eqnarray}
\frac{d{\cal B} \left ( B^+ \to \pi^+ \ell^+ \ell^- \right )} 
     {d q^2} 
& = & 
\frac{G_F^2 \alpha_{\rm em}^2 \tau_{B^+}}{1024 \pi^5 m_B^3} \, 
      \left | V_{tb} V_{td}^* \right |^2 \, 
\label{B-pi-ell-ell-HQS-limit} \\
& \times & 
      \sqrt{\lambda (q^2)} \sqrt{1 - \frac{4 m_\ell^2}{q^2}} \, 
      \tilde F (q^2) \, f_+^2 (q^2) , 
\nonumber 
\end{eqnarray}
%
where the dynamical function~$F (q^2)$, 
defined in Eq.~(\ref{eq:dynamical-function-def}), 
is now reduced to the following expression:   
\begin{eqnarray}
\tilde F (q^2) & = & 
\frac{2}{3} \, \lambda (q^2) 
\left ( 1 + \frac{2 m_\ell^2}{q^2} \right ) 
\left | 
C_9^{\rm eff} (q^2) + \frac{2 m_b}{m_B} \, C_7^{\rm eff} (q^2) 
\right |^2 
\nonumber \\
& + & 
\frac{2}{3} \, \lambda (q^2) 
\left |  C_{10}^{\rm eff} \right |^2  + 
\frac{4 m_\ell^2}{q^2} \left | C_{10}^{\rm eff} \right |^2 
\label{eq:Fq2-tilde} \\  
& \times &
\left [ 
\left ( 1 - \frac{m_\pi^2}{m_B^2} \right )^2 
\left ( m_B^2 - q^2 \right )^2 - 
\frac{2}{3} \, \lambda (q^2)  
\right ] , 
\nonumber   
\end{eqnarray}
and the kinematic function $\lambda (q^2)$ 
is given in Eq.~(\ref{eq:lambda}).

\begingroup 
\squeezetable 
\begin{table}[tb]
\caption{\label{tab:input}
Main input parameters used in the theoretical evaluations 
of the $B^+ \to \pi^+ \ell^+ \ell^-$ branching fractions 
taken from the PDG~\cite{Beringer:1900zz}, except for the 
$B$-meson leptonic decay constant~$f_B$, whose  value 
is taken from Lattice-NRQCD~\cite{Dowdall:2013tga}.    
} 
\begin{center}
\begin{tabular}{ll}
\hline 
$G_F = 1.11637 \times 10^{-5} \, \rm{GeV}^{-2}$ & 
$\alpha_{\rm em}^{-1} = 129$ \\  
$m_B = 5.2792 \,\rm{GeV}$ &
$\tau_{B^+} = 1.641$~ps \\
$m_\pi = 139.57 \,\rm{MeV}$ &
$f_\pi = 132$~MeV \\   
$\alpha_{s} (M_Z) = 0.1184 \pm 0.0007$ & 
$f_B = (184 \pm 4)$~GeV \\
$m_c (m_c) = (1.275 \pm 0.025)$~GeV &
$m_b (m_b) = (4.18 \pm 0.03)$~GeV \\  
$\lambda = 0.22535 \pm 0.00065$ & 
$A = 0.817 \pm 0.015$ \\
$\bar\rho = 0.136 \pm 0.018$ & 
$\bar\eta = 0.348 \pm 0.014$ \\
$|V_{ud}| = 0.97427$ & 
$|V_{tb}| = 0.999146$ \\
$|V_{ub}| = (3.51^{+0.15}_{-0.14}) \times 10^{-3}$ & 
$|V_{td}| = (8.67^{+0.29}_{-0.31}) \times 10^{-3}$ \\
   \hline
\end{tabular}
\end{center}
\end{table}
\endgroup  

Restricting ourselves to the NLL results for the 
effective Wilson coefficients (i.\,e., dropping 
the $\alpha_s (\mu)$-dependent terms in them) and 
using the $f_+ (q^2)$ form-factor shape extracted 
in terms of the BGL parametrization from the combined 
BaBar and Belle data, and the numerical values of the 
different quantities entering~(\ref{B-pi-ell-ell-HQS-limit}) 
from Table~\ref{tab:input}, the numerical values 
of the $B^\pm \to \pi^\pm \mu^+ \mu^-$ partial branching 
ratio in the ranges $4 m_\mu^2 \leq q^2 \leq 8$~GeV$^2$ and 
$1~{\rm GeV}^2 \leq q^2 \leq 8$~GeV$^2$ are given below:
%
\begin{equation}
\begin{split} 
{\cal B} (B^\pm \to \pi^\pm \mu^+ \mu^-; \, 
         0.05 \,{\rm GeV}^2 \leq q^2 \leq 8 \,{\rm GeV}^2) & \\
= (0.80 \pm 0.07) \times 10^{-8} , &  
\end{split} 
\label{eq:Br-0.05-8} 
\end{equation}
\begin{equation}
\begin{split} 
{\cal B} (B^\pm \to \pi^\pm \mu^+ \mu^-; \, 
         1 \,{\rm GeV}^2 \leq q^2 \leq 8 \,{\rm GeV}^2) & \\ 
= (0.72 \pm 0.06) \times 10^{-8} . & 
\end{split} 
\label{eq:Br-1-8}
\end{equation}

\subsection{Including HQS-Breaking Correction}
\label{ssec:with-SBCs}

Heavy-quark symmetry, which holds  
in the large-recoil limit, allows one to get relations among 
the $B \to \pi$ form factors~\cite{Charles:1998dr}. 
Taking into account the leading-order symmetry-breaking corrections, 
these relations were worked out in Ref.~\cite{Beneke:2000wa}:  
\begin{eqnarray}
f_0 (q^2) & = & 
\left ( 1 - \frac{q^2}{m_B^2} \right ) f_+ (q^2) 
\label{eq:f0-fp-rel} \\ 
& \times & 
\left \{ 1 + \frac{C_F \alpha_s (\mu_h)}{4\pi}  
\left [ 2 - 2  L (q^2) \right ] \right \} 
\nonumber \\
& + &
\frac{C_F \alpha_s (\mu_{hc})}{4\pi} \, 
\frac{q^2}{m_B^2 - q^2} \, \Delta F_\pi,
\nonumber 
\end{eqnarray}
\begin{eqnarray}
f_T (q^2) & = & 
\left ( 1 + \frac{m_\pi}{m_B} \right ) f_+ (q^2)
\label{eq:fT-fp-rel} \\ 
& \times & 
\left [ 1 + \frac{C_F \alpha_s (\mu_h)}{4\pi} 
\left ( \ln \frac{m_b^2}{\mu_h^2} + 2 L (q^2) \right ) \right ]  
\nonumber \\
& - & \frac{C_F \alpha_s (\mu_{hc})}{4\pi} \, 
\frac{m_B \left ( m_B + m_\pi \right )}{m_B^2 - q^2} \, \Delta F_\pi,
\nonumber 
\end{eqnarray}
where $C_F = 4/3$. The strong coupling~$\alpha_s (\mu)$ 
depends on the specific scales of the contributing diagrams, 
which we take as the hard $\mu_h \sim m_b$ and hard-collinear 
$\mu_{hc} \sim \sqrt{m_b \Lambda}$ scales, where 
$\Lambda \simeq 0.5$~GeV is the typical soft hadronic scale.  
The auxiliary function $L (q^2)$ is defined 
as follows~\cite{Beneke:2000wa}: 
\begin{equation}
L (q^2) = \left ( 1 - \frac{m_B^2}{q^2} \right ) 
\ln  \left ( 1 - \frac{q^2}{m_B^2} \right ) ,
\label{eq:L-function}
\end{equation}
with the normalization $L (0) = 1$,  
and the contributions of the hard-spectator diagrams 
are parametrized by the quantity~\cite{Beneke:2000wa}:   
\begin{equation}
\Delta F_\pi = \frac{8 \pi^2 f_B f_\pi}{3 m_B} 
\left\langle l_+^{-1} \right\rangle_+ 
\left\langle \bar u^{-1} \right\rangle_\pi .
\label{eq:DF-pi}
\end{equation}
Here,~$f_B$ and~$f_\pi$ are the leptonic decay constants 
of the $B$- and $\pi$-mesons, respectively, and the following first 
inverse moments of the $B$- and $\pi$-mesons are used: 

\begin{equation}
\left\langle l_+^{-1} \right\rangle_+ = 
\int\limits_0^\infty dl_+ \, \frac{\phi^B_+ (l_+)}{l_+} , 
\quad 
\left\langle \bar u^{-1} \right\rangle_\pi = 
\int\limits_0^1 du \, \frac{\phi_\pi (u)}{1 - u} , 
\end{equation}
which are completely determined by the leading-twist 
light-cone distribution amplitudes 
$\phi^B_+ (l_+)$~\cite{Grozin:1996pq,Braun:2003wx} 
and $\phi_\pi (u)$~\cite{Chernyak:1977as,Chernyak:1977fk,%
Chernyak:1980dj,Efremov:1979qk,Efremov:1978rn,Lepage:1979zb,%
Lepage:1980fj,Braun:1988qv,Ball:1998je}. With the input 
parameters~$m_B$, $f_B$ and~$f_\pi$ from Table~\ref{tab:input},  
and the  moments evaluated as  
$\left\langle \bar u^{-1} \right\rangle_\pi (1~{\rm GeV}) 
= 3.30 \pm 0.42$ and  
$\left\langle l_+^{-1} \right\rangle_+ (1.5~{\rm GeV}) 
= (1.86 \pm 0.17)$~GeV$^{-1}$~\cite{Lee:2005gza}, we estimate
 $\Delta F_\pi = 0.74 \pm 0.12$. 
This is numerically somewhat smaller than the value $\Delta F_\pi = 1.17$ 
used in Ref.~\cite{Beneke:2000wa}.  
This difference reflects the observation 
that the $\pi$-meson is well described 
by the asymptotic form of the twist-2 LCDA 
$\phi_\pi (u) = 6 u \left ( 1 - u \right )$, 
and the first subleading Gegenbauer moment 
$a_2 (1~{\rm GeV}) = 0.10 \pm 0.14$~\cite{Agaev:2012tm}
is consistent with zero. 

\begin{figure}[tb] 
\centerline{
\includegraphics[width = 0.45\textwidth]{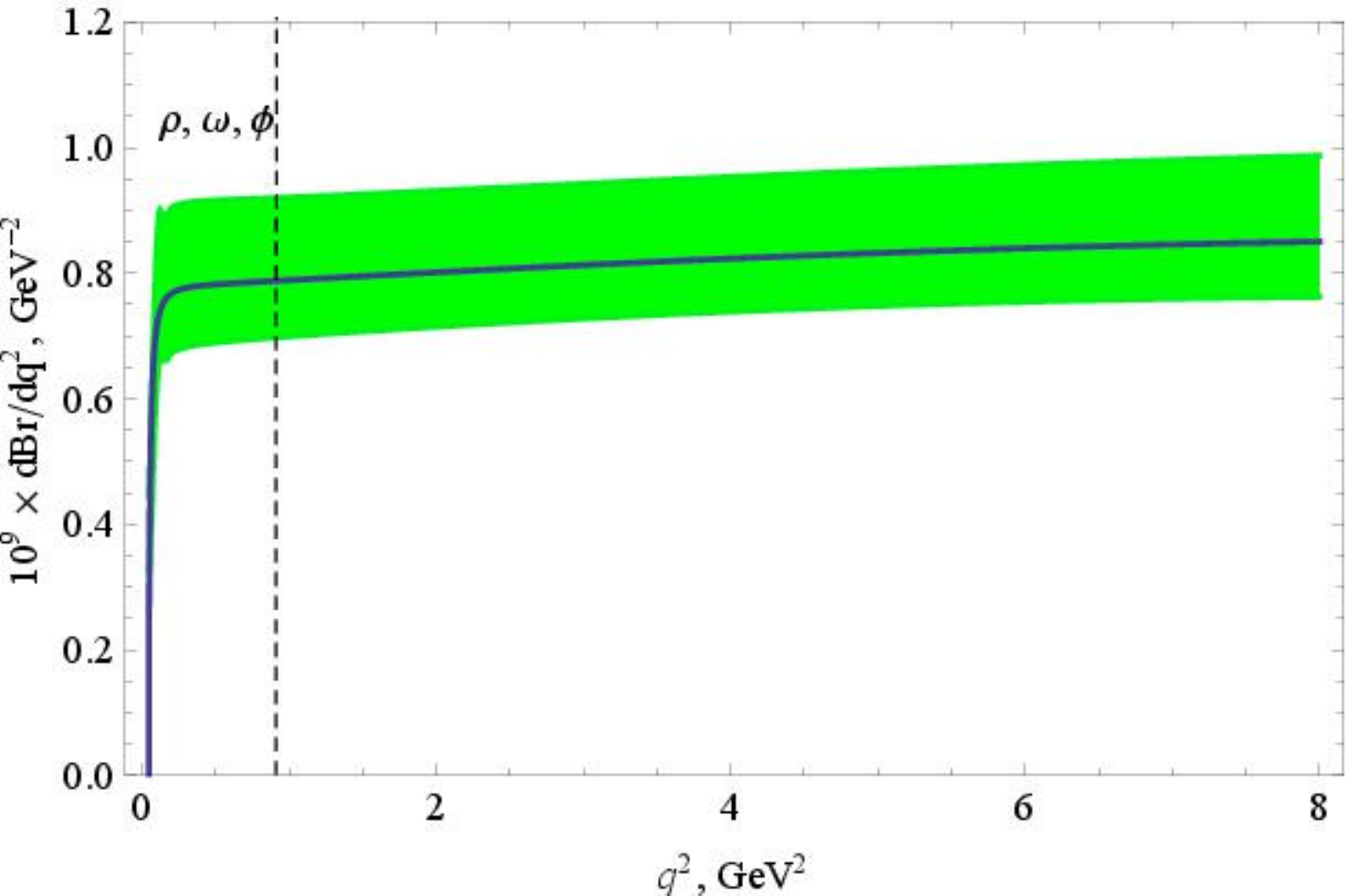}
} 
\caption{\label{fig:dB-dq2-1-8}
(Color online.)
The dilepton invariant-mass distribution 
$d{\cal B} (B^\pm \to \pi^\pm \ell^+ \ell^-)/dq^2$
for $0 \leq q^2 \leq 8$~GeV$^2$ calculated 
by taking into account the leading HQS-breaking corrections. 
Dashed vertical line indicates collectively the vector $\rho$-, 
$\omega$-, and $\phi$-resonance region.    
} 
\end{figure}

Taking into account the symmetry-breaking corrections, 
and the NNLO effects in the effective Wilson coefficients,
the partial branching fractions, integrated in the ranges 
of~$q^2$ as in Eqs.~(\ref{eq:Br-0.05-8}) and~(\ref{eq:Br-1-8}), 
are decreased. We get  
\begin{equation}
\begin{split}
{\cal B} (B^+ \to \pi^+ \mu^+ \mu^-; \, 
         0.05 \,{\rm GeV}^2 \leq q^2 \leq 8 \,{\rm GeV}^2) & \\
= (0.65^{+0.08}_{-0.06}) \times 10^{-8} , &  
\end{split}
\label{eq:Br-0.05-8-SBC} 
\end{equation}
\begin{eqnarray}
\begin{split}
{\cal B} (B^+ \to \pi^+ \mu^+ \mu^-; \, 
         1 \,{\rm GeV}^2 \leq q^2 \leq 8 \,{\rm GeV}^2) & \\
= (0.57^{+0.07}_{-0.05}) \times 10^{-8} , & 
\end{split}
\label{eq:Br-1-8-SBC}
\end{eqnarray}
which mainly reflects the NNLO effects in the Wilson coefficients. 
The corresponding dilepton invariant-mass distribution 
in the large-recoil approximation ($q^2 \leq 8$~GeV$^2$) 
is shown in Fig.~\ref{fig:dB-dq2-1-8}.
The vertical line shows the light-resonance 
($\rho$, $\omega$, and $\phi$) region collectively. 
The upper bound on~$q^2$ is imposed to avoid the large 
(resonant) contribution from the long-distance process 
$B^\pm \to \pi^\pm \, J/\psi \to \pi^\pm \ell^+ \ell^-$.

\begin{figure}[tb] 
\centerline{ 
\includegraphics[width = 0.45\textwidth]{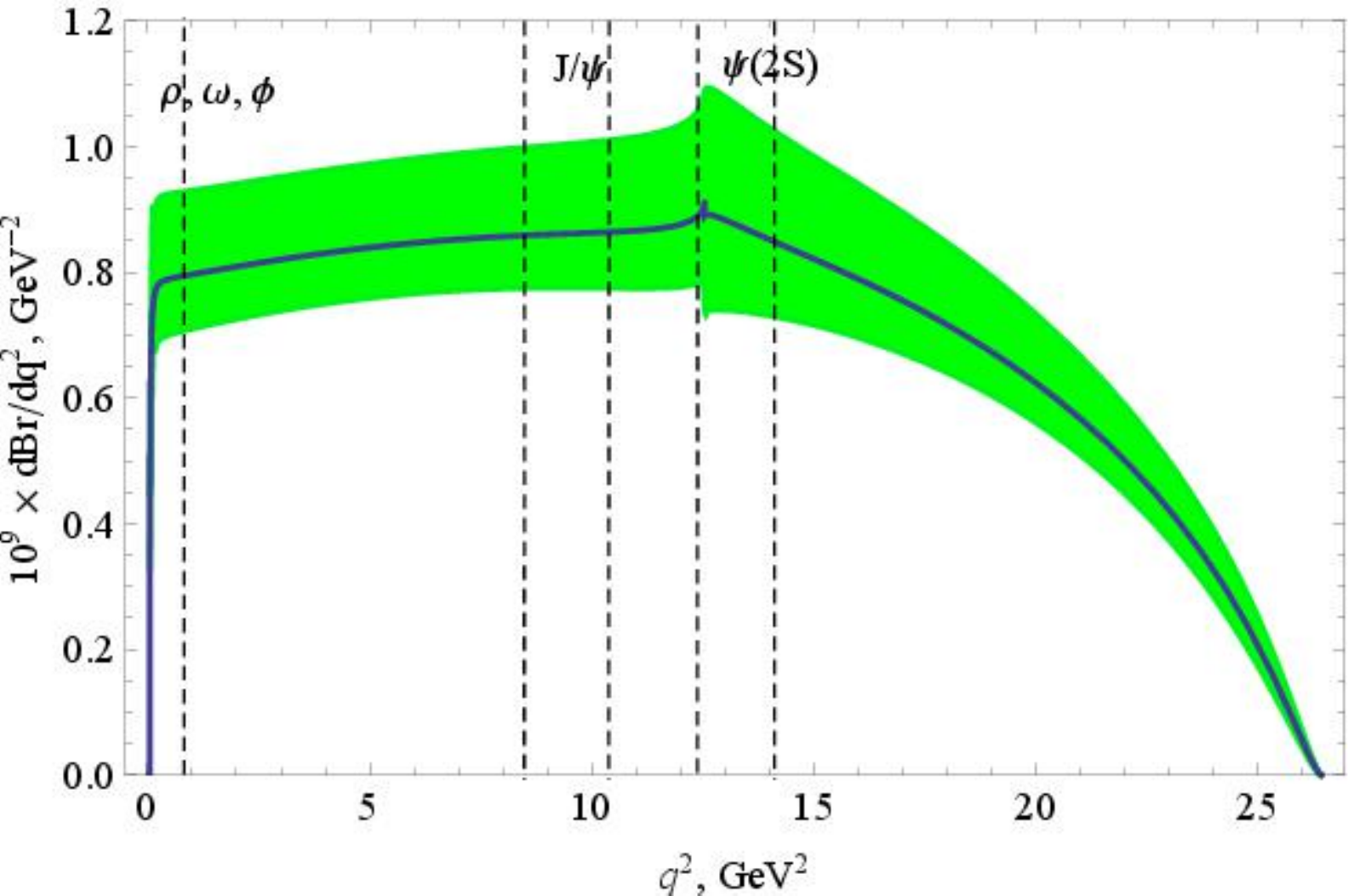} 
} 
\caption{\label{fig:dB-dq2-Entire} 
(Color online.)
The dilepton invariant-mass distribution 
in the $B^+ \to \pi^+ \ell^+ \ell^-$ decay 
for the entire kinematic range $0 \leq q^2 \leq 26.4$~GeV$^2$.
Dashed vertical lines specify the positions 
of vector resonances: $\rho$-, $\omega$- and $\phi$-mesons 
at $q^2 \lesssim 1$~GeV$^2$ and $J/\psi$- and $\psi(2S)$-mesons 
near $q^2 \simeq 9.5$~GeV$^2$ and $q^2 \simeq 13.5$~GeV$^2$, 
respectively. 
} 
\end{figure}

\begin{table}[tb]
\caption{\label{tab:dB-dq2-results}
Partial branching ratios $d{\cal B}(B^+ \to \pi^+ \mu^+ \mu^-)/dq^2$ 
integrated over the indicated ranges $[q^2_{\rm min}, q^2_{\rm max}]$.
}
%
%
\begin{center}
\begin{tabular}{|c|c|}
\hline 
$[q^2_{\rm min}, q^2_{\rm max}] $ & 
$10^8 \times {\cal B} (q^2_{\rm min} \le q^2 \le q^2_{\rm max})$ \\
\hline
$[ 0.05, 2.0  ] $ & $0.15^{+0.03}_{-0.02}$ \\[1mm]
$[ 1   , 2.0  ] $ & $0.08^{+0.01}_{-0.01}$ \\[1mm]
$[ 2.0 , 4.3  ] $ & $0.19^{+0.03}_{-0.02}$ \\[1mm]
$[ 4.3 , 8.68 ] $ & $0.37^{+0.06}_{-0.04}$ \\[1mm]
$[10.09, 12.86] $ & $0.25^{+0.04}_{-0.03}$ \\[1mm]
$[14.18, 16.0 ] $ & $0.15^{+0.03}_{-0.02}$ \\[1mm]
$[16.0 , 18.0 ] $ & $0.15^{+0.03}_{-0.02}$ \\[1mm]
$[18.0 , 22.0 ] $ & $0.25^{+0.04}_{-0.03}$ \\[1mm]
$[22.0 , 26.4 ] $ & $0.13^{+0.02}_{-0.02}$ \\[1mm]
\hline
$[  0.05, 8.0 ] $ & $0.66^{+0.10}_{-0.07}$ \\[1mm]
$[  1.0 , 8.0 ] $ & $0.58^{+0.09}_{-0.06}$ \\[1mm]
\hline
$[4 m_\mu^2, (m_B - m_\pi)^2] $ (total) 
& $1.88 ^{+0.32}_{-0.21}$ \\
\hline
\end{tabular}
\end{center}
\end{table}

\subsection{\label{ssec:ansatz-accuracy}  
Estimating the $SU (3)_F$-Breaking \\ 
in the $B \to \pi, K$ Tensor Form Factors 
} 

Before presenting the estimates  
of the $B^+ \to \pi^+ \ell^+ \ell^-$ branching fraction 
in the entire kinematic range of~$q^2$, we would like to
discuss  the validity of the Ansatz~(\ref{eq:FFT-ratio-def})
used by us in calculating the $SU(3)_F$-breaking effects 
in the  $B \to \pi, K$ tensor form factors. 
The accuracy of our Ansatz can be easily determined 
in the kinematic region where the HQS-based methods apply. 
These will be worked out below and used to project also 
the accuracy in the large-$q^2$ region. We note that 
the lattice data already provides a reliable estimate 
of the r.h.s. of Eq.~(\ref{eq:FFT-ratio-def}), but only 
preliminary lattice data~\cite{Bouchard:2013zda} 
are available for the l.h.s., involving the tensor 
form factors.

Taking into account the leading-order HQS-symmetry-breaking 
effects, all three $B \to P$ transition form 
factors, where~$P$ is a light pseudoscalar meson, 
are related, as shown in  Eqs.~(\ref{eq:f0-fp-rel}) 
and~(\ref{eq:fT-fp-rel}). This then allows one to relate 
the $SU (3)_F$-symmetry breaking measures: 
\begin{eqnarray}
&& 
R_0 (q^2) = R_+ (q^2) 
\left [ 1 + \frac{C_F \alpha_s (\mu_{hc})}{4\pi} \, 
\right. 
\label{eq:R0-Rp-rel} \\  
&& \hspace*{15mm} 
\left. \times  
\frac{q^2/m_B^2}{\left (1 - q^2/m_B^2 \right )^2} \, 
\left ( \frac{\Delta F_K}{f_+^{BK} (q^2)} - 
        \frac{\Delta F_\pi}{f_+^{B\pi} (q^2)} \right ) 
\right ] , 
\nonumber   
\end{eqnarray}
\begin{eqnarray}
&& 
R_T (q^2) = 
\frac{1 + m_K/m_B}{1 + m_\pi/m_B} \, R_+ (q^2) 
\left [ 1 - \frac{C_F \alpha_s (\mu_{hc})}{4\pi} 
\qquad  
\right. 
\label{eq:RT-Rp-rel} \\ 
&& \hspace*{15mm} 
\left. \times 
\frac{1}{1 - q^2/m_B^2} \, 
\left ( \frac{\Delta F_K}{f_+^{BK} (q^2)} - 
        \frac{\Delta F_\pi}{f_+^{B\pi} (q^2)} \right ) 
\right ] , 
\nonumber 
\end{eqnarray}
where  
\begin{equation} 
f_+^{BK} (q^2) = f_+^{B\pi} (q^2) 
\left [ 1 + R_+ (q^2) \right ] ,
\label{eq:fpBK-fpBpi}
\end{equation}
and  
\begin{eqnarray} 
\Delta F_K & = & \Delta F_\pi \, \frac{f_K}{f_\pi} \, 
\frac{\langle \bar u^{-1} \rangle_K}
     {\langle \bar u^{-1} \rangle_\pi} 
\label{eq:DFK-DFpi} \\ 
& \simeq & 
\Delta F_\pi \left ( 1 + \Delta f_{K\pi} \right ) 
\left [ 1 + a_1^K (\mu_{hc}) \right ] . 
\nonumber 
\end{eqnarray}
Here $\Delta f_{K\pi} = f_K/f_\pi - 1 \simeq 0.23$  
is the $SU(3)_F$-symmetry breaking in the leptonic 
decay constants ($f_\pi \simeq 130$~MeV and 
$f_K \simeq 160$~MeV~\cite{Beringer:1900zz}). 
The first inverse moments of the $K$- and $\pi$-mesons 
$\langle \bar u^{-1} \rangle_P (\mu_{hc}) \simeq 
3 \left [ 1 + a_1^P (\mu_{hc}) \right ]$ are 
approximated by the asymptotic and the first 
Gegenbauer terms in the conformal expansion 
of the LCDAs with $a_1^\pi (2~{\rm GeV}) = 0$ and 
$a_1^K (2~{\rm GeV}) = 0.05 \pm 0.02$~\cite{Ball:2006fz,Ball:2007zt}  
(the other terms in the Gegenbauer decomposition do not 
affect the ratio $\Delta F_K/\Delta F_\pi$ significantly).  
Keeping terms linear in~$\Delta f_{K\pi}$, $a_1^K (\mu_{hc})$ 
and~$R_+ (q^2)$ only in the hard-collinear 
correction, the measures of the $SU(3)_F$-symmetry 
breaking become: 
\begin{eqnarray}
&& 
R_0 (q^2) = R_+ (q^2) 
\left \{ 
1 + \frac{C_F \alpha_s (\mu_{hc})}{4\pi} \, 
\right. 
\label{eq:R0-Rp-rel-approx} \\  
&& \hspace*{5mm}
\left. \times 
\frac{\hat q^2 \, \Delta F_\pi}
     {\left (1 - \hat q^2 \right )^2 f_+^{B\pi} (q^2)}  
\left [ \Delta f_{K\pi} + a_1^K (\mu_{hc}) - R_+ (q^2) \right ] 
\right \} , 
\nonumber 
\end{eqnarray}
\begin{eqnarray}
&& 
R_T (q^2) = 
\frac{1 + \hat m_K}{1 + \hat m_\pi} \, R_+ (q^2) 
\left \{ 
1 - \frac{C_F \alpha_s (\mu_{hc})}{4\pi} \, 
\right. 
\label{eq:RT-Rp-rel-approx} \\ 
&& \hspace*{5mm}
\left. \times 
\frac{\Delta F_\pi}
     {\left ( 1 - \hat q^2 \right ) f_+^{B\pi} (q^2)}  
\left [ \Delta f_{K\pi} + a_1^K (\mu_{hc}) - R_+ (q^2) \right ] 
\right \} , 
\nonumber 
\end{eqnarray}
where the reduced mass $\hat m_P = m_P/m_B$ ($P = \pi, \, K$) 
and the reduced momentum transfer squared
is defined as $\hat q^2 = q^2/m_B^2$. 
With $\hat m_\pi = 0.0265$ 
and $\hat m_K = 0.0947$~\cite{Beringer:1900zz}, 
their difference $\hat m_K - \hat m_\pi = 0.0682$ 
yields $(1 + \hat m_K)/(1 + \hat m_\pi) = 1.07$ 
for the prefactor on the r.h.s. 
of Eq.~(\ref{eq:RT-Rp-rel-approx}).

To quantify the validity 
of the Ansatz~(\ref{eq:FFT-ratio-def}),  
let us introduce the following function: 
\begin{equation}
\Delta R (q^2) = \frac{1}{2} 
\left [ R_+ (q^2) + R_0 (q^2) \right ] - R_T (q^2) ,   
\label{eq:DR-def}
\end{equation}
whose deviation from zero quantitatively determines the 
accuracy of our $SU(3)_F$-breaking Ansatz. Using Eqs.~(\ref{eq:R0-Rp-rel-approx})
and~(\ref{eq:RT-Rp-rel-approx}), $\Delta R (q^2)$ can be 
estimated as follows: 
\begin{eqnarray}
&& 
\Delta R (q^2) \simeq R_+ (q^2) 
\left \{ \hat m_\pi - \hat m_K + 
\frac{C_F \alpha_s (\mu_{hc})}{4\pi} \, 
\right. 
\label{eq:DR-approx} \\  
&& \hspace*{3mm}
\left. \times 
\frac{\left ( 1 - \hat q^2/2 \right ) \Delta F_\pi}
     {\left (1 - \hat q^2 \right )^2 f_+^{B\pi} (q^2)}   
\left [ \Delta f_{K\pi} + a_1^K (\mu_{hc}) - R_+ (q^2) \right ] 
\right \} .  
\nonumber 
\end{eqnarray}
There are two competitive contributions: 
the first one is coming from the reduced mass 
difference, and 
the second one combines the perturbative corrections 
in the form factors (the HQS-breaking corrections 
due to the hard-spectator contributions). 

To remove the term induced by the $K$- and 
$\pi$-meson difference from $\Delta R (q^2)$, 
we define a reduced function  $\tilde R_T (q^2)$ as follows:  
\begin{equation}
\tilde R_T (q^2) \equiv
\frac{m_B + m_\pi}{m_B + m_K} \, 
\frac{f_T^{BK} (q^2)}{f_T^{B\pi} (q^2)} - 1 ,  
\label{eq:FF-ratio-mod}
\end{equation}
and a reduced analogue  
of the $\Delta R (q^2)$ function: 
\begin{equation}
\Delta \tilde R (q^2) \equiv \frac{1}{2} 
\left [ R_+ (q^2) + R_0 (q^2) \right ] - 
\tilde R_T (q^2) .    
\label{eq:DR-mod-def}
\end{equation}
In the low-$q^2$ region 
(say, $0 \le q^2 \le 14$~GeV$^2$ 
or $0 \le \hat q^2 \le 1/2$),
the deviation of this function 
from zero is completely determined by the 
hard-spectator corrections in the form factors:  
\begin{eqnarray}
&& 
\Delta \tilde R (q^2) \simeq R_+ (q^2) \, 
\frac{C_F \alpha_s (\mu_{hc})}{4\pi} \, 
\label{eq:DR-mod-approx} \\  
&& \hspace*{7mm} 
\times
\frac{\left ( 1 - \hat q^2/2 \right ) \Delta F_\pi}
     {\left (1 - \hat q^2 \right )^2 f_+^{B\pi} (q^2)}   
\left [ \Delta f_{K\pi} + a_1^K (\mu_{hc}) - R_+ (q^2) \right ] . 
\nonumber 
\end{eqnarray}
The input parameters are as follows: $C_F = 4/3$, 
the hard-collinear scale $\mu_{hc} = 2$~GeV, 
$\alpha_s (m_\tau) = 0.330 \pm 0.014$~\cite{Beringer:1900zz}, 
$\left ( 1 - \hat q^2 \right ) f_+^{B\pi} (q^2) \simeq 
f_+^{B\pi} (0) = 0.260 \pm 0.014$ (our estimate), 
$\Delta F_\pi = 0.74 \pm 0.12$ (our estimate), 
$\Delta f_{K\pi} = f_K/f_\pi - 1 = 0.23$~\cite{Beringer:1900zz}, 
$a_1^K (2~{\rm GeV}) = 0.05 \pm 0.02$~\cite{Ball:2006fz,Ball:2007zt}.
In addition, we need to know $R_+(q^2)$. Ignoring the mild $q^2$-dependence,
we set $R_+(q^2)\simeq R_+(0)$, and discuss some representative
estimates of $R_+(0)$.  
The most recent lattice result for the $B \to K$ vector form 
factor is by the HPQCD 
Collaboration~\cite{Bouchard:2013mia,Bouchard:2013eph}
 $f_+^{BK} (0) = 0.319 \pm 0.066$. With the
determination of the corresponding quantity in the $B \to \pi$
transition,  $f_+^{B\pi} (0) = 0.260 \pm 0.014$, 
we get $R_+ (0) = 0.231 \pm 0.262$ (the error 
is dominated by the uncertainty in $f_+^{BK} (0)$). 
Another recent estimate 
$f_+^{BK} (0) = 0.33 \pm 0.04$~\cite{Becirevic:2012fy}  
yields  $R_+ (0) = 0.269 \pm 0.169$, where 
again the error is mainly due to $f_+^{BK} (0)$. 
Note that the LCSR estimate  
$f_+^{BK} (0) = 0.34^{+0.05}_{-0.02}$~\cite{Khodjamirian:2010vf}  
is compatible with the above lattice   
predictions within the uncertainties. 
After the insertion of the lattice estimates 
in Eq.~(\ref{eq:DR-mod-approx}), the results 
are as follows: 
\begin{equation}  
\Delta \tilde R (q^2) \simeq 
\frac{1 - \hat q^2/2}{1 - \hat q^2} 
\left \{ 
\begin{array}{ll} 
(1.15 \pm 2.19) \times 10^{-3}, & 
\text{\cite{Bouchard:2013mia,Bouchard:2013eph}} \\ 
(0.31 \pm 1.58) \times 10^{-3} . & 
\text{\cite{Becirevic:2012fy}} 
\end{array} 
\right. 
\label{eq:DR-mod-numerics}
\end{equation} 
So, the effect of the hard-scattering corrections 
is below~1\% in the kinematic domain considered. 

Coming back to the numerical evaluation 
of $\Delta R (q^2)$, defined in~(\ref{eq:DR-def}), 
using the estimates~(\ref{eq:DR-mod-numerics}) 
given above, one obtains: 
\begin{eqnarray}  
\Delta R (q^2) & \simeq & R_+ (q^2) 
\left ( \hat m_\pi - \hat m_K \right ) 
\label{eq:DR-numerics} \\ 
& \simeq & 
\left \{ 
\begin{array}{ll} 
(- 1.55 \pm 1.76) \times 10^{-2} , & 
\text{\cite{Bouchard:2013mia,Bouchard:2013eph}} \\ 
(- 1.81 \pm 1.13) \times 10^{-2} . & 
\text{\cite{Becirevic:2012fy}} 
\end{array} 
\right. 
\nonumber 
\end{eqnarray} 
So, the uncertainty of the Ansatz~(\ref{eq:FFT-ratio-def}) 
can be evaluated to be approximately~3\% 
in the considered range of~$q^2$.

The estimates presented above support the 
Ansatz~(\ref{eq:FFT-ratio-def}) within an accuracy 
of about~3\%. To which degree of accuracy, this 
Ansatz also holds in the high-$q^2$ domain will be tested 
as the lattice calculations of all the $B \to \pi$ 
transition form factors become completely quantitative.
We include an additional error of~5\%, ascribed to 
the error on the Ansatz~(\ref{eq:FFT-ratio-def}) in the 
determination of the tensor form factor $f_T^{B\pi}(q^2)$.

\section{\label{sec:entire-region}  
$B^+ \to \pi^+ \ell^+ \ell^-$ Decay in the Entire $q^2$-Range
} 

In the low hadronic-recoil region (large-$q^2$),  
heavy-quark symmetry does not hold, and we have 
three independent form factors $f_+ (q^2)$, $f_0 (q^2)$ 
and $f_T (q^2)$ in $B^\pm \to \pi^\pm \ell^+\ell^-$. 
We have given a detailed account of their determination 
in the preceding sections. The vector form factor $f_+(q^2)$ 
is determined taking into account the Belle 
and BaBar data on $B \to \pi \ell \nu_\ell$, and fitting several
parametrizations, with the BGL-parametrization as our default choice.
We have used the HQS-based method, including the 
leading-order symmetry breaking, in the low-$q^2$ region ($q^2 \leq 8$~GeV$^2$),
and the experimentally constrained form factor $f_+(q^2)$ to
determine the other two form factors $f_0(q^2)$ and $f_T(q^2)$.
Finally, we have used the available 
Lattice-QCD results for the form factors $f_i^{BP}(q^2)$ $(i=+,0,T)$
in the large-$q^2$ region, 
obtained for the $B \to K$ and $B \to \pi$ transitions. As the lattice data on
$f_T^{B\pi}(q^2)$ is still sparse, we have determined this form factor
from the lattice data on $f_T^{BK}(q^2)$, and an Ansatz for the $SU(3)_F$-breaking.
We have tested the accuracy of this Ansatz in the low-$q^2$ region, and find it to hold
within 3\%. This dedicated study has removed the largest source of theoretical uncertainty
originating from the form factors. 

Before presenting our numerical results, we discuss 
the choice for the parameter $\sqrt z = m_c/m_b$ 
entering the NNLO corrections. The NNLO corrections 
to the $b \to s \, \ell^+ \ell^-$ transition matrix 
element~\cite{Greub:2008cy}, which we have adapted 
for the exclusive $b \to d \, \ell^+ \ell^-$ case 
discussed by us here, are available in the literature both 
as the Mathematica and the C++ programs~\cite{Greub:2008cy}, 
from which the former one was implemented in our own 
Mathematica routine. We need to fix this ratio in terms 
of the $c$- and $b$-quark pole masses. The three-loop 
relation between the pole $m_{\rm pole}$ and 
$\overline{\rm MS}$-scheme $\bar m (\bar m)$ 
masses~\cite{Chetyrkin:1999ys,Chetyrkin:1999qi,Melnikov:2000qh} 
can be used to get the $c$- and $b$-quark pole masses.   
Staring from the values collected in Table~\ref{tab:input}, 
the ratio $m_c (m_c)/m_b (m_b) = 0.305 \pm 0.006$ 
can be transformed into the ratio of the pole masses 
$m_{c, {\rm pole}}/m_{b, {\rm pole}} = 0.402 \pm 0.008$.     
In~\cite{Xing:2007fb}, additional electroweak 
corrections to the relation between the pole and the
$\overline{\rm MS}$ quark masses were taken into account with 
the resulting pole masses: $m_{c, {\rm pole}} = 1.77 \pm 0.14$~GeV 
and $m_{b, {\rm pole}} = 4.91 \pm 0.12$~GeV, with the 
ratio $m_{c, {\rm pole}}/m_{b, {\rm pole}} = 0.36 \pm 0.03$. 
This value is used by us as input for~$\sqrt z$ 
in calculating the $c$-quark loop-induced corrections.

The invariant-mass spectrum in the entire range  
of~$q^2$ ($4 m_\ell^2 < q^2 < 26.4$~GeV$^2$) 
is presented in Fig.~\ref{fig:dB-dq2-Entire}. 
Once again, we emphasize that this represents only the
short-distance contribution.
The dashed vertical lines specify the light-meson resonant 
region, shown at $q^2 \lesssim 1$~GeV$^2$, 
as well as of the $J/\psi$- and $\psi(2S)$-mesons.   
In the calculation of 
this spectrum, Wilson coefficients 
are used in the NNLO accuracy. In the perturbative improvement, the
auxiliary functions $F_{1,2}^{(7)} (q^2)$  
and $F_{1,2}^{(9)} (q^2)$ entering the 
next-to-leading correction in $C_9^{\rm eff} (q^2)$  
are known analytically as  power expansions 
in~$s = q^2/m_B^2$ and in~$1 - s$ 
(as shown in Fig.~\ref{fig:F-12-79-corrections}). As explained earlier,
we have extrapolated these functions 
into the intermediate $q^2$-region. In doing this, 
we have matched the known analytical functions 
in the form of expansions at the 
``matching'' point $q^2 \simeq 12.5$~GeV$^2$, 
at which value the spectrum has the minimal 
discontinuity (see Fig.~\ref{fig:dB-dq2-Entire}).
This yields an invariant-mass spectrum which 
is a smooth function of~$q^2$, within uncertainties.
It is important to note that the ``matching'' 
point $q^2 \simeq 12.5$~GeV$^2$ lies in 
the $\psi (2S)$-resonance region which is 
dominated by the long-distance effects. Away from the
resonance regions, the short-distance 
contribution to the differential branching 
fraction dominates and the discontinuity 
in the spectrum discussed earlier is not a crucial issue.

Our predictions for the partial branching fractions 
$d{\cal B} (B^\pm \to \pi^\pm \, \ell^+ \ell^-)/dq^2$ 
in eleven different~$q^2$ bins are presented 
in Table~\ref{tab:dB-dq2-results}. 
The total branching fraction of the semileptonic 
$B^\pm \to \pi^\pm \, \mu^+ \mu^-$ decay is as follows: 
%
\begin{equation} 
\begin{split}  
& 
{\cal B} (B^\pm \to \pi^\pm \, \mu^+ \mu^-) \\ 
& \quad = 
\left ( 
1.88 ^{+0.28}_{-0.15} \big |_{\mu_b} \pm 0.13 \big |_{|V_{td}|} 
     \pm 0.08 \big |_{\rm FF} \pm 0.01 
\right ) \times 10^{-8} \\
& \hspace*{10mm} =  
\left ( 1.88 ^{+0.32}_{-0.21} \right ) \times 10^{-8} , 
\end{split}  
\label{eq:BF-total}
\end{equation} 
%
where the individual uncertainties are from 
the scale dependence~$\mu_b$ of the Wilson coefficients, 
the CKM matrix element~$|V_{td}|$, and the form factors (FF), as
indicated. The resulting average uncertainty is about~15\%, 
which is dominated by the scale dependence of the Wilson 
coefficients and can be reduced after the scale-dependence 
of the tensor form factor $f^{B\pi}_T (q^2)$ is worked out 
properly in the entire $q^2$-range.   

The branching fraction for the semileptonic 
$B^\pm \to \pi^\pm \, e^+ e^-$ decay is the same 
as~(\ref{eq:BF-total}), as the additional contribution 
induced by the shift to the lower kinematic values 
of $q^2 = 4 m_e^2 \simeq 1$~MeV$^2$ is negligible. 

The use of the isospin symmetry allows to make 
predictions for the $B^0 \to \pi^0 \, \ell^+ \ell^-$  
decay also. Neglecting the effects of the isospin 
symmetry breaking in the $B \to \pi$ transition 
form factors which are expected to be  a few 
percent, the main modification is the isospin 
factor $C_{\pi^0} = 1/2$ in the final state due 
to the $\pi^0$-meson structure. Taking this into 
account, our predictions for the partial branching 
fractions are as follows:   
\begin{equation}
\begin{split}
{\cal B} (B^0 \to \pi^0 \ell^+ \ell^-; \, 
         0.05 \,{\rm GeV}^2 \leq q^2 \leq 8 \,{\rm GeV}^2) & \\
= (0.33^{+0.05}_{-0.03}) \times 10^{-8} , &  
\end{split}
\label{eq:Br-0.05-8-SBC-0} 
\end{equation}
\begin{eqnarray}
\begin{split}
{\cal B} (B^0 \to \pi^0 \ell^+ \ell^-; \, 
         1 \,{\rm GeV}^2 \leq q^2 \leq 8 \,{\rm GeV}^2) & \\
= (0.29^{+0.05}_{-0.03}) \times 10^{-8} , & 
\end{split}
\label{eq:Br-1-8-SBC-0}
\end{eqnarray}
where $\ell = e$ or~$\mu$, and for the total 
branching fraction we estimate:  
\begin{equation} 
{\cal B} (B^0 \to \pi^0 \, \ell^+ \ell^-) =  
\left ( 0.94^{+0.16}_{-0.11} \right ) \times 10^{-8} .  
\label{eq:BF-total-0}
\end{equation} 
%
The above decay rates ${\cal B} (B^0 \to \pi^0 \, \ell^+ \ell^-)$ will  be measured 
at the forthcoming Super-B factory at KEK.


\section{\label{sec:summary} 
Summary and Outlook
} 

We have presented a theoretically improved 
calculation of the branching fraction for 
the $B^\pm \to \pi^\pm \mu^+ \mu^-$ decay, 
measured recently by the LHCb Collaboration~\cite{LHCb:2012de}. 
In doing this, we have used the effective 
Wilson coefficients $C_7^{\rm eff} (q^2)$, 
$C_9^{\rm eff} (q^2)$ and $C_{10}^{\rm eff}$, 
obtained in the NNLO accuracy earlier for 
the $b \to (s, \, d) \, \ell^+ \ell^-$ 
decays~\cite{Bobeth:1999mk,Asatrian:2001de,%
Asatryan:2001zw,Ali:2002jg,Asatrian:2003vq}. 
Some of the auxiliary functions, called 
$F_{1,2}^{(7)} (q^2)$, $F_{1,2}^{(9)} (q^2)$,  
$F_{1,(2),u}^{(7)} (q^2)$, $F_{1,(2),u}^{(9)} (q^2)$
are known analytically in the limiting case
of $m_c/m_b = 0$~\cite{Seidel:2004jh}, which 
we have used. For realistic values of this ratio,
taken by us as $\sqrt z = m_c/m_b = 0.36$, 
the results are known only in limited ranges of 
$s = q^2/m_B^2$ ($s \leq 0.35$ and $0.55 < s < 1.0$). 
All these functions are shown numerically 
in Fig.~\ref{fig:F-12-79-corrections}. 
We have interpolated in the gap, which introduces 
some uncertainty, but being part of the NNLO contribution, 
it is not expected to be the dominant error. Theoretical
uncertainties are dominated by the imprecise knowledge 
of the form factors, $f_+^{B\pi} (q^2)$ and $f_T^{B\pi} (q^2)$. 
We have extracted the shape of the former from data on 
the charged-current process $B \to \pi \ell \nu_\ell$, 
measured at the $B$-factories.  
Among the four popular parametrizations, the BGL 
(Boyd, Grinstein and Lebed) 
$z$-expansion was chosen as our working tool.  
For the tensor form factor $f_T^{B\pi} (q^2)$, 
heavy-quark symmetry provides the information 
in the low-$q^2$ (large-recoil) region, in which 
this form factor is related to the known factor 
$f_+^{B\pi} (q^2)$, up to symmetry-breaking effects, 
which we have estimated from the existing literature. 
This provides us an estimate of the dilepton 
invariant-mass spectrum for $q^2 \leq 8$~GeV$^2$.  
For larger values of~$q^2$, we have used the 
$SU(3)_F$-symmetry-breaking Ansatz and knowledge 
of the form factor $f_T^{BK} (q^2)$
from lattice QCD. Comparison with the 
preliminary results by the HPQCD Collaboration studies 
of the form factor $f_T^{B\pi}(q^2)$ in the low-recoil 
(or large-$q^2$) region~\cite{Bouchard:2013zda} shows 
a good consistency with our results. This then provides 
us a trustworthy profile of the two form factors needed 
in estimating the entire dilepton invariant-mass spectrum 
and the partial branching ratio. The combined accuracy 
on the branching ratio is estimated as~$\pm 15\%$, 
and the resulting branching fraction 
${\cal B} (B^\pm \to \pi^\pm \mu^+ \mu^-) = 
(1.88 ^{+0.32}_{-0.21})\times 10^{-8}$ 
is in agreement with the LHCb data~\cite{LHCb:2012de}. 
We have provided partial branching fractions in
different ranges of~$q^2$, which can be compared directly 
with the data, as and when they become available.

\bigskip 

\textit{Note added in Proofs.}  
Recently, the analysis of the $B \to \pi \ell \bar \ell$ 
and $B \to \pi \rho \bar \ell$ decays in the relativistic 
quark model has been presented in Ref.~\cite{Faustov:2014zva}. 
The main difference in comparison with our analysis is that 
the $B \to \pi$ transition form factors were determined 
theoretically by utilizing the relativistic quark model 
based on the quasipotential approach and QCD. The total 
branching fraction ${\cal B} (B^\pm \to \pi^\pm \mu^+ \mu^-) = 
(2.0 \pm 0.2) \times 10^{-8}$ is in good agreement with 
our result.


\begin{acknowledgments}
A.~R. would like to thank Wei Wang and Christian Hambrock 
for helpful discussions on technical details of the calculations
performed and the Theory Group at DESY for the kind and generous 
hospitality. We acknowledge helpful communications with Ran Zhou 
of the FermiLab and MILC Collaborations on the lattice results.  
We are thankful Aoife Bharucha, Alexander Khodjamirian and 
Yu-Ming Wang for their comments on the vector form factor 
and long-distance effects. 
The work of A.~R. is partially supported by the German-Russian
Interdisciplinary Science Center (G-RISC) funded by the German 
Federal Foreign Office via the German Academic Exchange Service 
(DAAD) under the project No. P-2013a-9. 
\end{acknowledgments}


\bibliography{APR-PRD}

\end{document}